\documentclass{appolb}
\usepackage{epsfig}
\begin{document}
\title{Few-Nucleon Calculations and Correlations.
\thanks{Presented at 5th workshop on "e.-m. induced two-hadron emission";
Lund, Sweden, June 13-16, 2001.}
}
\author{W.~Gl\"ockle, H.~Kamada, J.~Golak, A.~Nogga
\address{Institut f\"ur Theoretische Physik II, Ruhr-Universit\"at Bochum,
D44780 Bochum, Germany}
\and
H.~Wita{\l}a, R.~Skibi\'nski, J.~Kuro\'s-\.Zo{\l}nierczuk
\address{Institute of Physics, Jagellonian University, PL-30059 Cracow,Poland}
}
\maketitle
\begin{abstract}
 Present day results for few-nucleon bound state and scattering
observables based on modern  high precision nuclear  forces
are briefly reviewed.
While
in relation to  NN forces of that type
three-nucleon (3N) forces
are mandatory for binding
energies and  for quite a few 3N scattering observables their effect is rather
small in two-nucleon correlation functions as demonstrated for $^3$He and
$^4$He.
 
The old idea of the Coulomb sum rule as a way to extract the $pp$
correlation function is reconsidered and the need for more accurate data
is pointed out. It appears to be an ideal case to probe properties of the
density operator and the ground state wave functions without disturbances
of final state interactions (FSI).
 
In the 3N system below the pion threshold
FSI is well under control and therefore the
exclusive process $^3{\rm He}(e,e'{\rm NN})$ is also a very good test case for
correlated nuclear wave functions and electromagnetic current operators.
One specific kinematics is emphasized, which can lead to insights
into the correlated ground state wave functions with little disturbance
of  FSI.
Finally exclusive photodisintegration of $^3$He is regarded, which
appears to be promising to  identify 3N force effects.
\end{abstract}
\PACS{21.45+v, 21.10-k, 25.10+s, 25.20-x, 27.10+h}
  
\section{Introduction}

The first and simplest description of nuclear physics is based
on the non-relativistic Schr\"odinger equation. On this stage
one requires that NN interactions are well tuned to NN data
up to the pion threshold and are accompanied by 3N forces, which at least
guarantee the correct $^3$H binding energy. For electromagnetic probes
current operators consistent with the nuclear interactions are needed.
To get insight into the dynamics of the nuclear systems, reliable
solutions of such a Schr\"odinger equation are necessary.
But extensions to this picture are possible.
A more advanced and possibly necessary
 dynamical picture would include relativity for instance in the instant
 form of a Hamiltonian formalism\cite{ref1}.
Here we restrict ourselves to a strictly
 nonrelativistic treatment. We employ the present day perfectly well tuned
 NN forces  CD-Bonn\cite{ref2}, AV18\cite{ref3},  Nijmegen I and
II\cite{ref4}.
They are mostly of
 phenomenological and local  nature, with the exception of CD-Bonn, which is a
 slightly modified one-boson-exchange potential and highly nonlocal.
 As 3N forces we choose the Tucson-Melbourne (TM)
2$\pi$-exchange~\cite{ref5}
 and the Urbana IX~\cite{ref6} models. In the first case the off-shell $\pi -  N$
amplitude
 is used in a low momentum expansion, in the second case that amplitude
 is based on an intermediate $\Delta$-excitation and a phenomenological short
 range part. Two strengths parameters in the Urbana IX 3N force are
 adjusted to the $^3$H binding energy and nuclear matter  density. In the
 TM 3N force we adjust the cut-off parameter $\Lambda$   for a strong form-factor
 parametrization separately for each NN force partner to the $^3$H
binding energy\cite{ref7}.
The 3N system is solved rigorously in the Faddeev scheme for bound and
 scattering states; the 4N bound state is equally precisely evaluated
 using Yakubovsky equations\cite{ref7}. A survey of typical and current results
 for binding energies and scattering observables is presented in Section~II.
 
Wave function properties in form of bound state two-body correlation
 functions are shown in Section~III. One approach of connecting them to
 observables is the Coulomb sum rule. We briefly review that topic in
 Section~IV and point to necessary
improvements in experiment and theory in order
 to achieve clear and convincing results in the future. 

Another approach in investigating correlations are electron induced
 two-nucleon emissions, which we study in a specific kinematics for
 the target nucleus $^3$He. It is shown in Section~V that FSI
appear to be  unavoidable (at least below the pion
threshold) but can possibly be reduced to an easily accessible and
restricted one. This might enable a search for initial state
 correlations in a rather controlled manner.
 
Photon induced two-nucleon emission on $^3$He as well as the  $pd$ break-up appear
 to be very promising to see 3N force effects. This is illustrated in
 Section~VI. Finally we end with a brief outlook.
                      
\section{Three-and Four-Nucleon Systems}

\begin{figure}
\begin{center}
%\psfrag{xxx}{\Large $E$($^3$H) [MeV] }
%\psfrag{yyy}{\Large $E$($^4$He) [MeV] }
\psfig{file=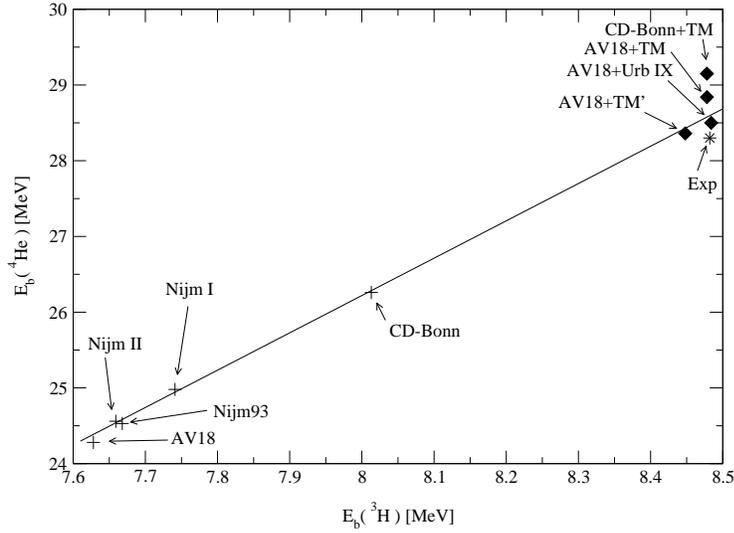,angle=-90,scale=0.4}
\end{center}
\caption{Tjon-line: $\alpha$-particle binding energy predictions $E$($^4$He)
         against the predictions for the $^3$H binding energy
         for several  interaction models.
         Predictions  without (crosses) and with (diamonds)  3N forces
         are shown. The experimental point is marked by a star. The line
         represents a least square fit to NN force predictions only.
}
\label{fig:tjonline}
\end{figure}

A first necessary test of the dynamical picture are three-and four-nucleon
 binding energies. For five modern, high precision  NN forces our theoretical results
 are shown in Table~\ref{tab:trialphabindnn}.
We have taken into account charge independence and charge
symmetry breaking as well as the mass difference of the proton
and the neutron. Further electromagnetic interactions are included and
also the isospin $T={3 \over 2}$ admixtures.
We see the by now well known under binding against
 the experimental values. Table~\ref{tab:trialphabind3nf} collects the
 individually adjusted
   $\Lambda$-parameters of the TM 3N force and of a modified one (TM'),
which violates chiral invariance less than TM, and the resulting
3N and 4N binding energies.
We also show the AV18 plus Urbana IX results.
We
 end up with the interesting result that the theoretical $\alpha$-particle
binding energies  for the different NN and 3N force combinations are rather
 close to the experimental value. There is a slight over binding, which
 leaves little room for the action of 4N forces. This can be quantified in
 the following way. The average attraction due to NN forces is 24.92 MeV
 or 88 \%    from
  28.30 MeV. The average additional attraction due to 3N forces is 3.9 MeV or
14 \%
 and finally the average over binding is 0.5 MeV, which is 2 \% of 28.30 MeV.
 If as a conjecture this would be attributed to  a repulsive 4NF then this
 shows a nice hierarchy in the importance of two- to many-body forces.
 Of course this is a temporary statement and can be modified in the future
 if more will be known about strengths and properties of 3N forces.
 In any case
 such a hierarchy is in agreement with the expectations of chiral
 perturbation theory\cite{ref8,ref9}.
An overview  of 3N and 4N  binding energies is shown in
 Fig.~\ref{fig:tjonline},
 which documents the strong correlation among them,
 known as Tjon line \cite{ref7,ref7a}.
 
\begin{table}                                 
 
%\parbox[t]{7.5cm}
{  \begin{center}
    \begin{tabular}{l|rrr}
 Potentials    &  $E$($^3$H) &$E$($^3$He)& $E$($^4$He) \\[2pt]
\hline
 & & & \\[-4pt]
Nijm~93        & -7.668      & -7.014    & -24.53      \\[2pt]
Nijm~I         & -7.741      & -7.083    & -24.98      \\[2pt]
Nijm~II        & -7.659      & -7.008    & -24.56      \\[2pt]
AV18           & -7.628      & -6.917    & -24.28      \\[2pt]
CD-Bonn        & -8.013      & -7.288    & -26.26      \\[2pt]
\hline
 & & & \\[-4pt]
Exp.           & -8.482      & -7.718    & -28.30      \\[2pt]
    \end{tabular}
 
    \caption{$^3$H, $^3$He and $^4$He binding energy predictions
             for several NN potential models compared to the experimental
values.
             All energies are given in MeV. }
    \label{tab:trialphabindnn}
  \end{center}}
%\hspace{0.5cm}
\end{table}
\begin{table} 
%\parbox[t]{7.5cm}
{  \begin{center}
    \begin{tabular}{l|r|rrr}
 Potentials    &  $\Lambda$ [$m_\pi$]  & $E$($^3$H)&$E$($^3$He)&
                                                        $E$($^4$He) \\[2pt]
\hline
 & & & \\[-4pt]
CD-Bonn+TM     & 4.784       & -8.478    & -7.735    & -29.15  \\[2pt]
AV18+TM        & 5.156       & -8.478    & -7.733    & -28.84  \\[2pt]
AV18+TM'       & 4.756       & -8.448    & -7.706    & -28.36  \\[2pt]
AV18+Urbana IX & ---         & -8.484    & -7.739    & -28.50  \\[2pt]
\hline
 & & & \\[-4pt]
Exp.           & ---         & -8.482    & -7.718    & -28.30  \\[2pt]
    \end{tabular}
 
    \caption{$^3$H, $^3$He and $^4$He binding energy predictions
             for several NN and 3N potential models compared to the experimental values.
             All energies are given in MeV.}
    \label{tab:trialphabind3nf}
  \end{center}}
 
\end{table}
 
%%%%%%%%  FIGURE 1 

%%%%%%%%  FIGURE 2 
\begin{figure}
\begin{center}
\psfig{file=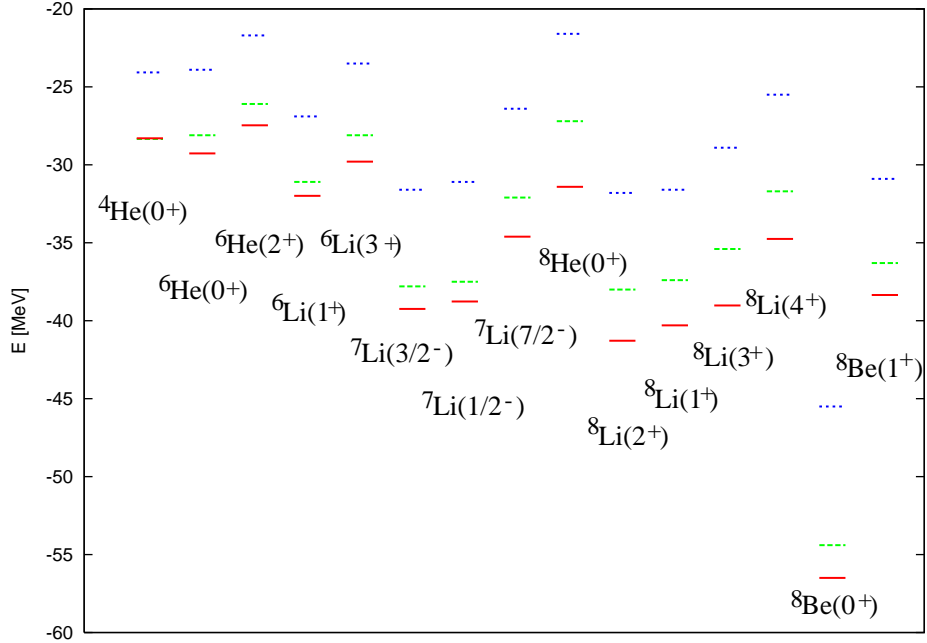,angle=-90,scale=0.5}
\end{center}
\caption{
  Spectra of light nuclei: experimental data (red,solid),
      AV18 (blue,dotted), AV18+Urbana~IX (green,dashed).
}
\label{fig:spec2}
\end{figure}

%%%%%%%%  FIGURE 3 
\begin{figure}[htb]
\begin{center}
\psfig{file=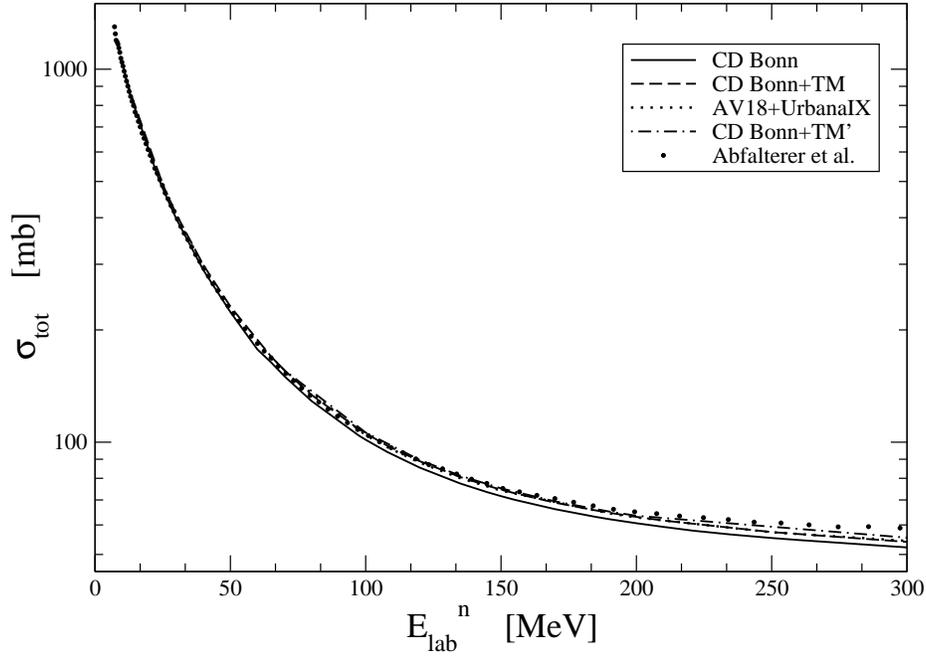,angle=-90,scale=0.5}
\end{center}
\caption{The total $nd$ cross section. Comparison of the data~\cite{ref30}
with various interaction models.}
\label{fig:ndtot}
\end{figure}
 
%%%%%%%%  FIGURE 4 
\begin{figure}[htb]
  \begin{center}
    \parbox[b]{0.5\linewidth}{
      \includegraphics[width=0.9\linewidth]{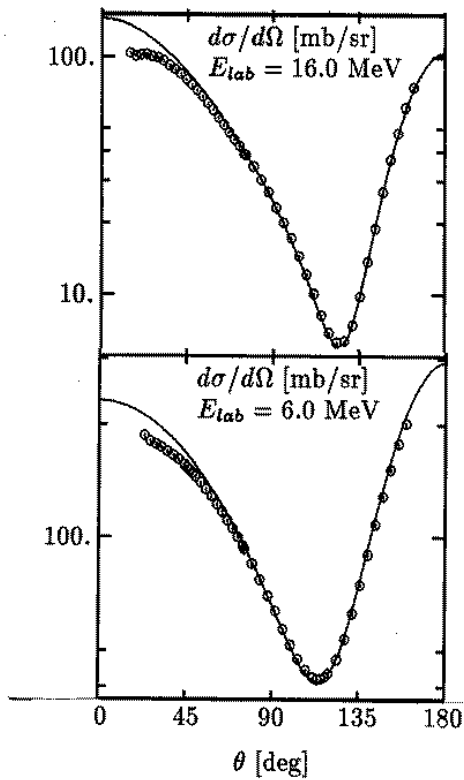}}
    \parbox[b]{7mm}{~}
    \parbox[b]{0.3\linewidth}{
      \caption[experimental setup]{\label{fig:fig9n}
        Differential cross section for elastic nucleon--deuteron
scattering. $pd$ data from \cite{ref16}.   }
      }
  \end{center}
\end{figure}

%%%%%%%%  FIGURE 5 
\begin{figure}[htb]
\begin{center}
\psfig{file=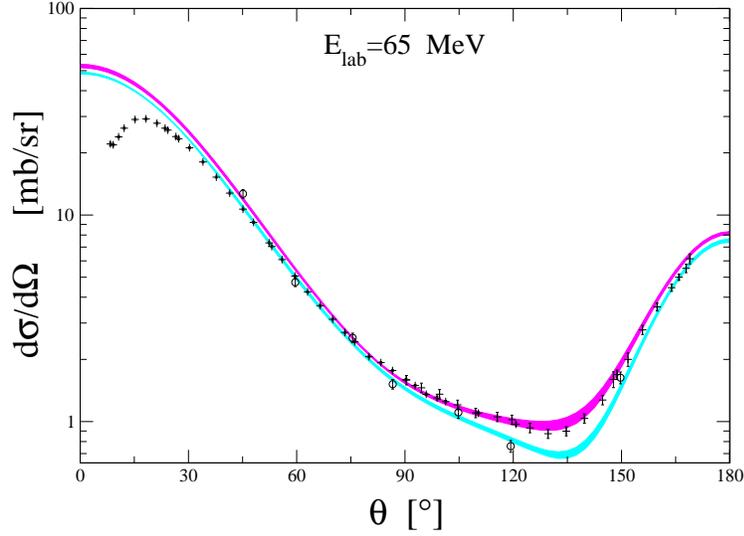,angle=-90,scale=0.4}
\end{center}
\caption{Differential cross section in elastic $Nd$ scattering at 65 MeV.
The light (dark) shaded bands are NN force only (NN+3NF) predictions for
various interactions. $pd$ data (crosses) are from \cite{shimizu} and $nd$
data (open circles) are from \cite{ruehl}.}
\label{fig:e65p.elastic.ds}
\end{figure}              
 
%%%%%%%%  FIGURE 6 
\begin{figure}[htb]
\begin{center}
\psfig{file=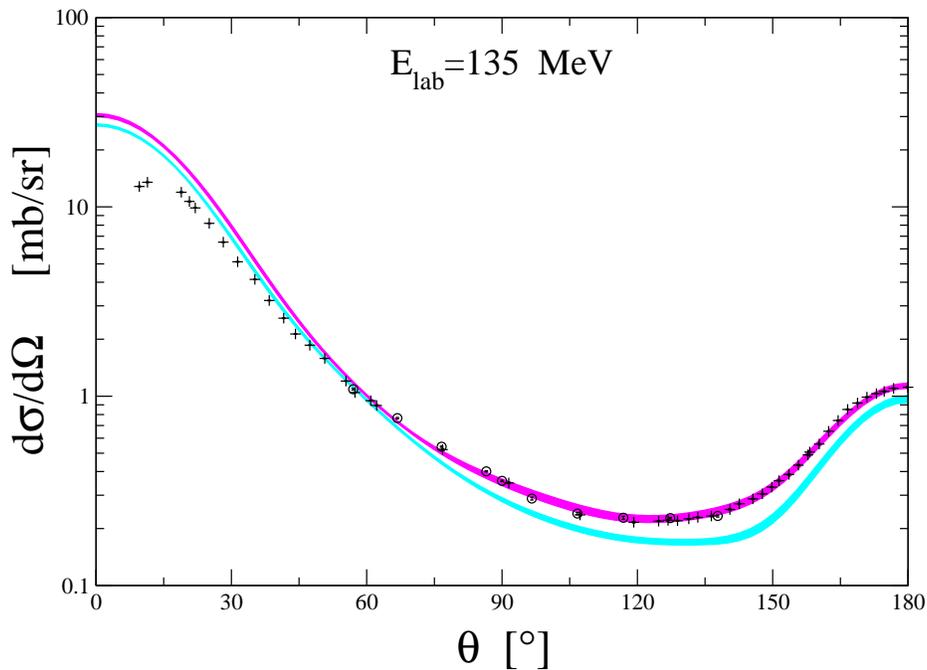,angle=-90,scale=0.5}
\end{center}
\caption{The same as in Fig.~\ref{fig:e65p.elastic.ds} for E= 135 MeV.
$pd$ data (crosses) are from \cite{ref17} and (circles) from \cite{ref18}.}
\label{fig:e135.elastic.ds}
\end{figure}
 
%%%%%%%%  FIGURE 7 
\begin{figure}[htb]
  \begin{center}
    \parbox[b]{0.5\linewidth}{
     \includegraphics[width=0.9\linewidth]{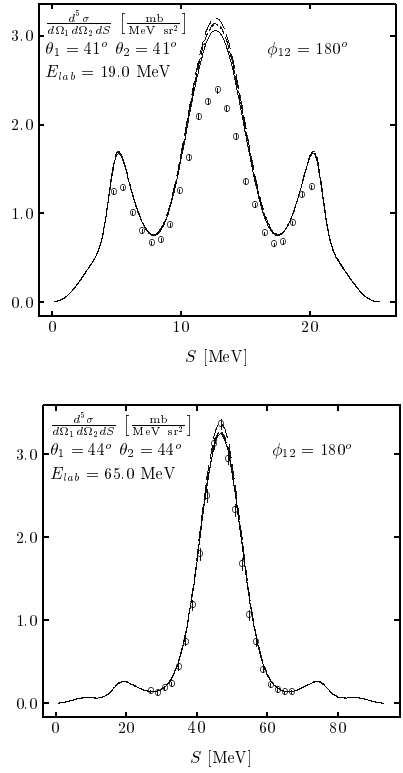}}
    \parbox[b]{7mm}{~}
    \parbox[b]{0.3\linewidth}{
      \caption[experimental setup]{\label{fig:fig35n}
       Five-fold differential $Nd$ breakup cross section along the
kinematical locus including the quasi-free scattering condition.
$pd$ data at 19 MeV are from ref.~\cite{ref19}
and at 65 MeV from ref.~\cite{ref20}.
      }
      }
  \end{center}
\end{figure}
 
%%%%%%%%  FIGURE 8                                                 
\begin{figure}
\begin{center}
\psfig{file=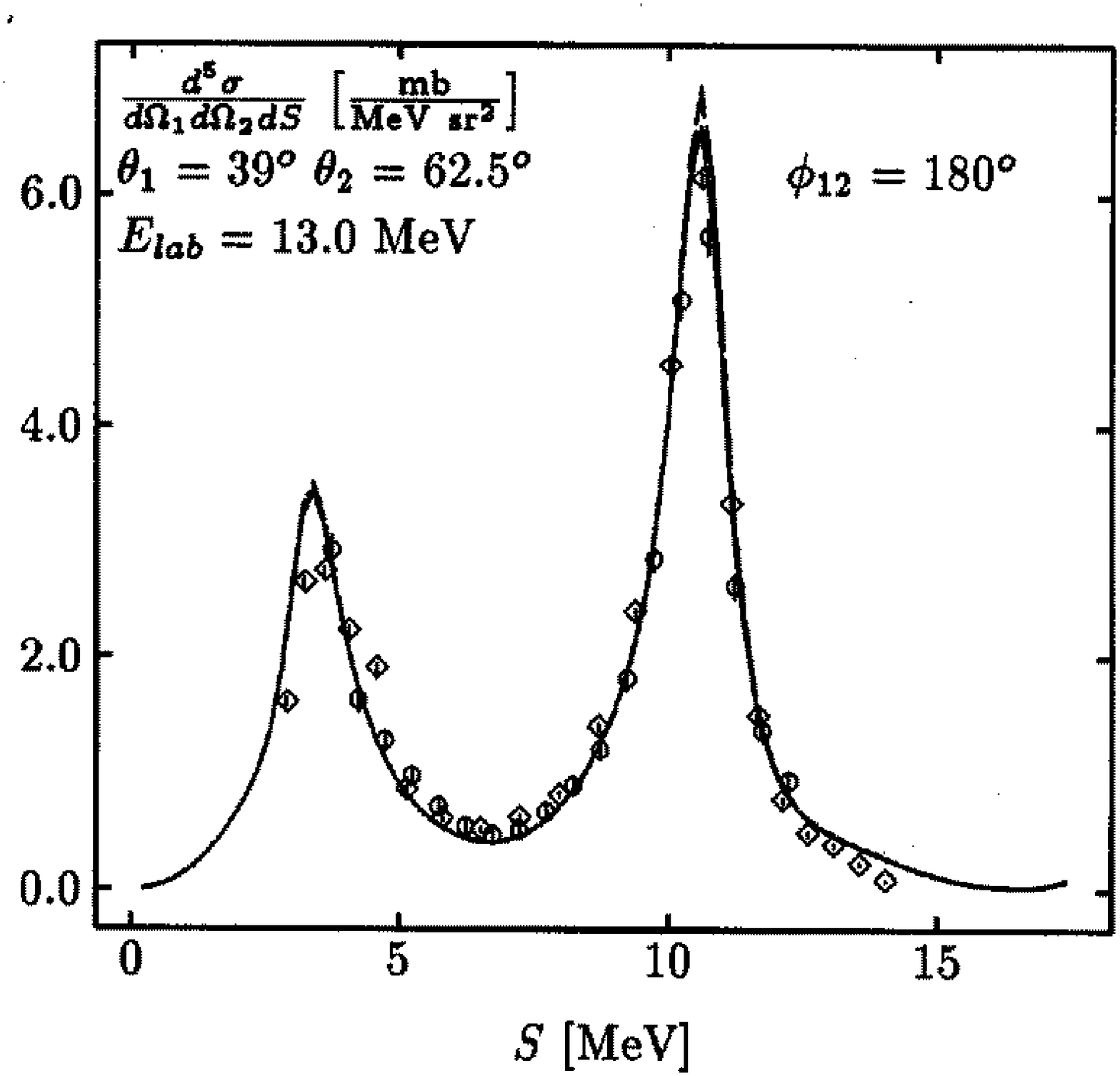,angle=0,scale=0.25}
\psfig{file=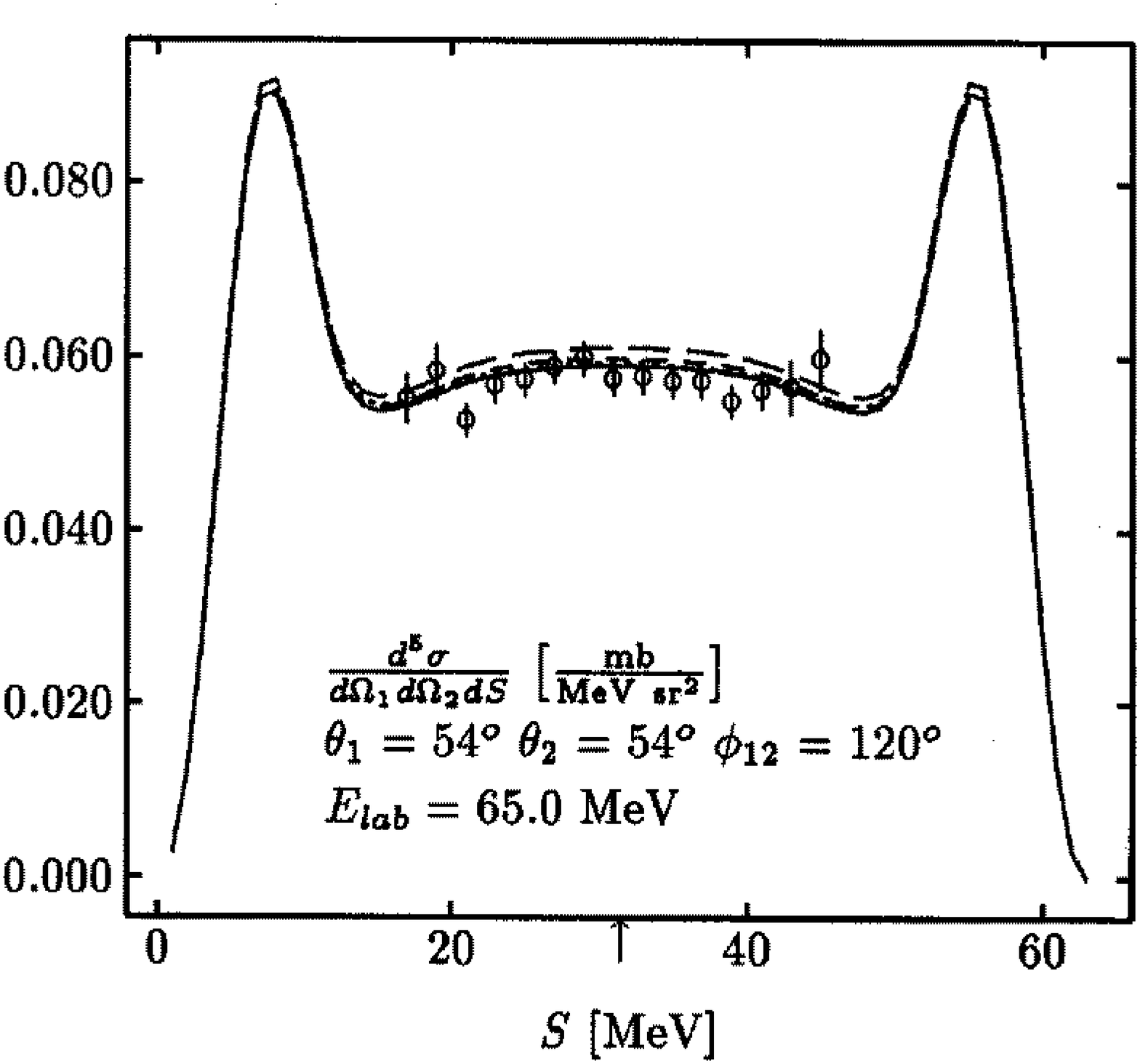,angle=0,scale=0.25}
\end{center}
\caption{The same as in Fig.~\ref{fig:fig35n} for two cases:
       two nucleons leave with equal momenta (up) and three nucleons
leave with equal energies under pairwise angles of 120$^\circ$ (down).
$pd$ data at 13 MeV are from ref.~\cite{ref21}, at 65 MeV from
ref.~\cite{ref22}.}
\label{fig:fig39n}
\end{figure}

%%%%%%%%  FIGURE 9 
\begin{figure}
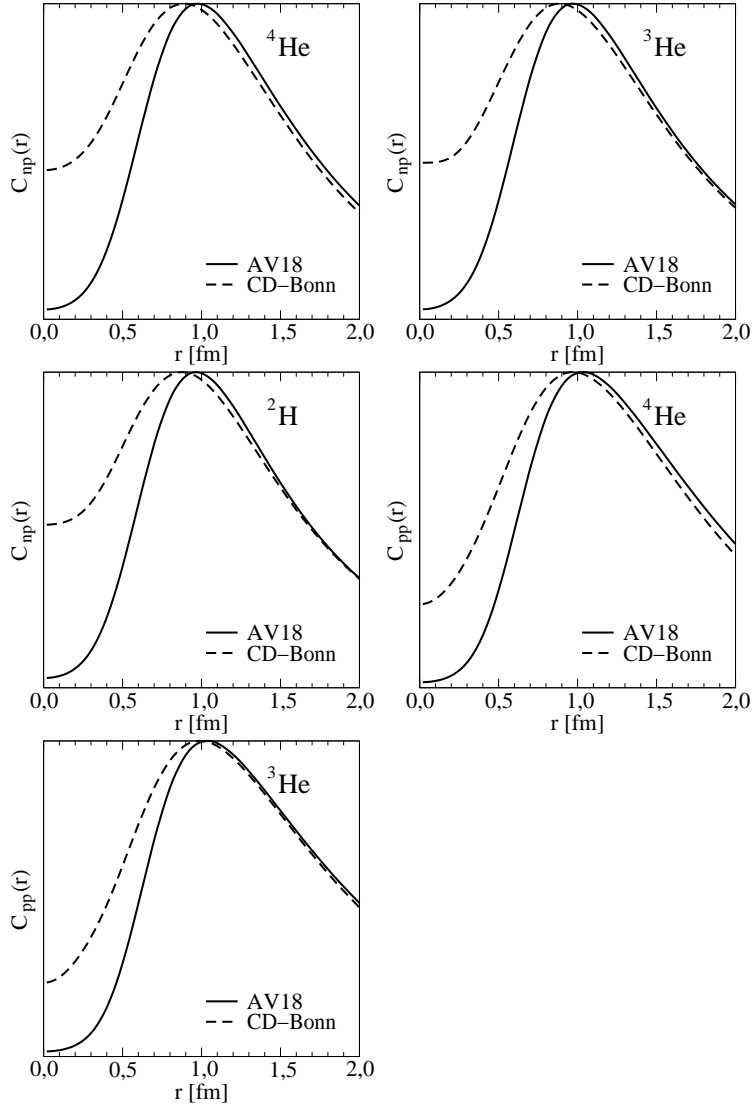

\begin{center}
\psfig{file=he4np.eps,angle=0,scale=0.5}
\psfig{file=he3np.eps,angle=0,scale=0.5}
\psfig{file=deutnp.eps,angle=0,scale=0.5}
\psfig{file=he4pp.eps,angle=0,scale=0.5}
\psfig{file=he3pp.eps,angle=0,scale=0.5}
\phantom{\psfig{file=deutnp.eps,angle=0,scale=0.5}}
\end{center}
\caption{Two-nucleon correlation functions $C(r)$ for $np$ and $pp$ pairs in
the nuclei $d$, $^3$He and $^4$He based on the AV18 and CD Bonn NN
potentials.}
\label{fig:crnppp}
\end{figure}

%%%%%%%%  FIGURE 10 
\begin{figure}
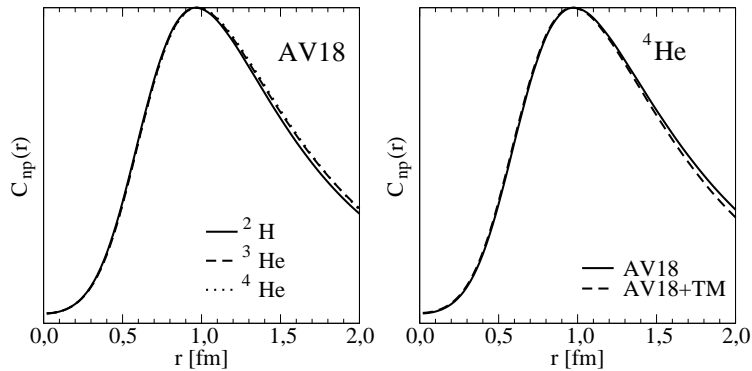

\begin{center}
\psfig{file=d3he4henp.eps,angle=0,scale=0.5}
\psfig{file=he4np3nf.eps,angle=0,scale=0.5}
\end{center}
\caption{Comparison of $C(r)$'s for $d$, $^3$He and $^4$He (left)
and for AV18 against AV18+TM (right).}
\label{fig:cr3nf}
\end{figure}
 
%%%%%%%%  FIGURE 11  
\begin{figure}
\begin{center}
\psfig{file=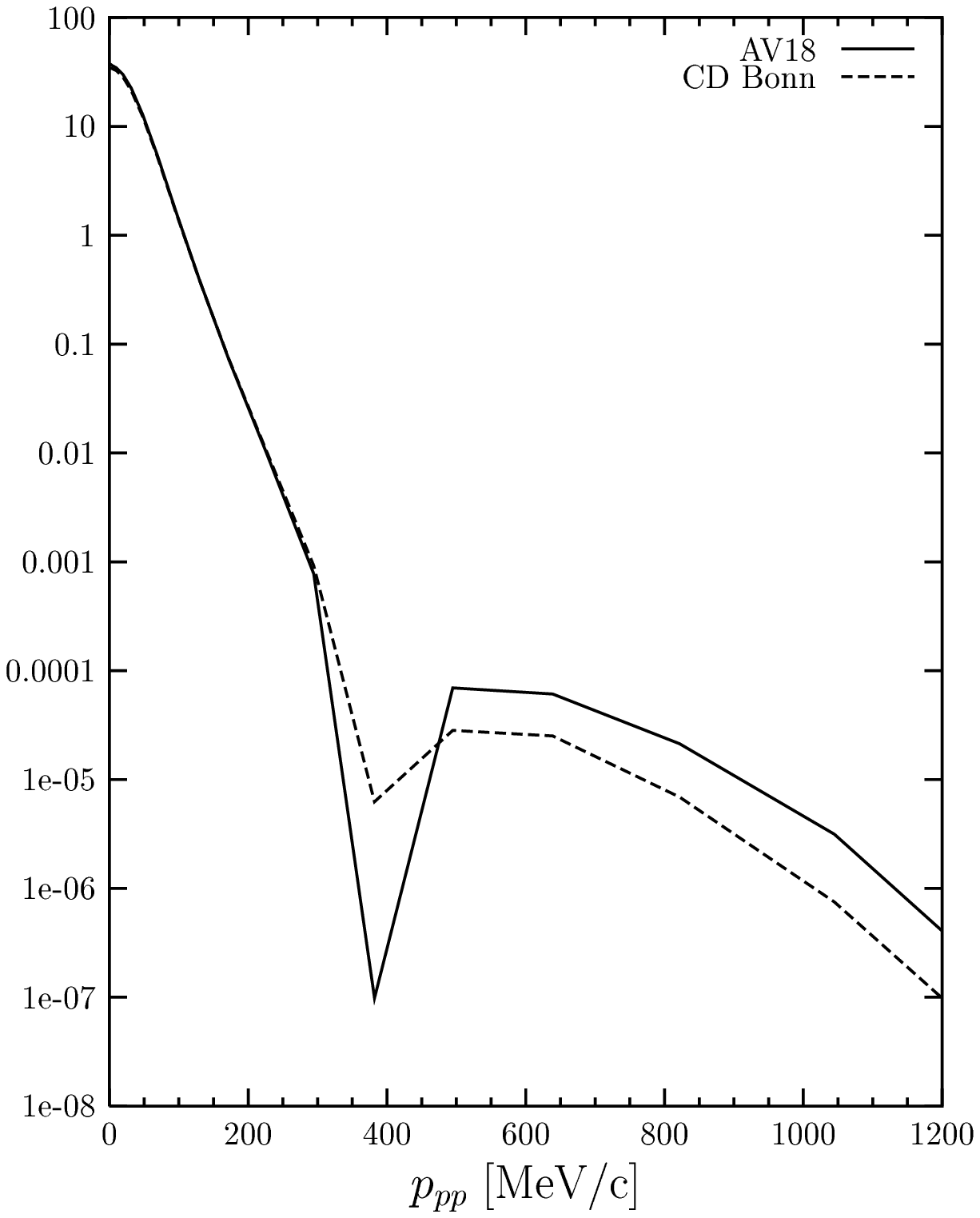,angle=0,scale=0.4}
\psfig{file=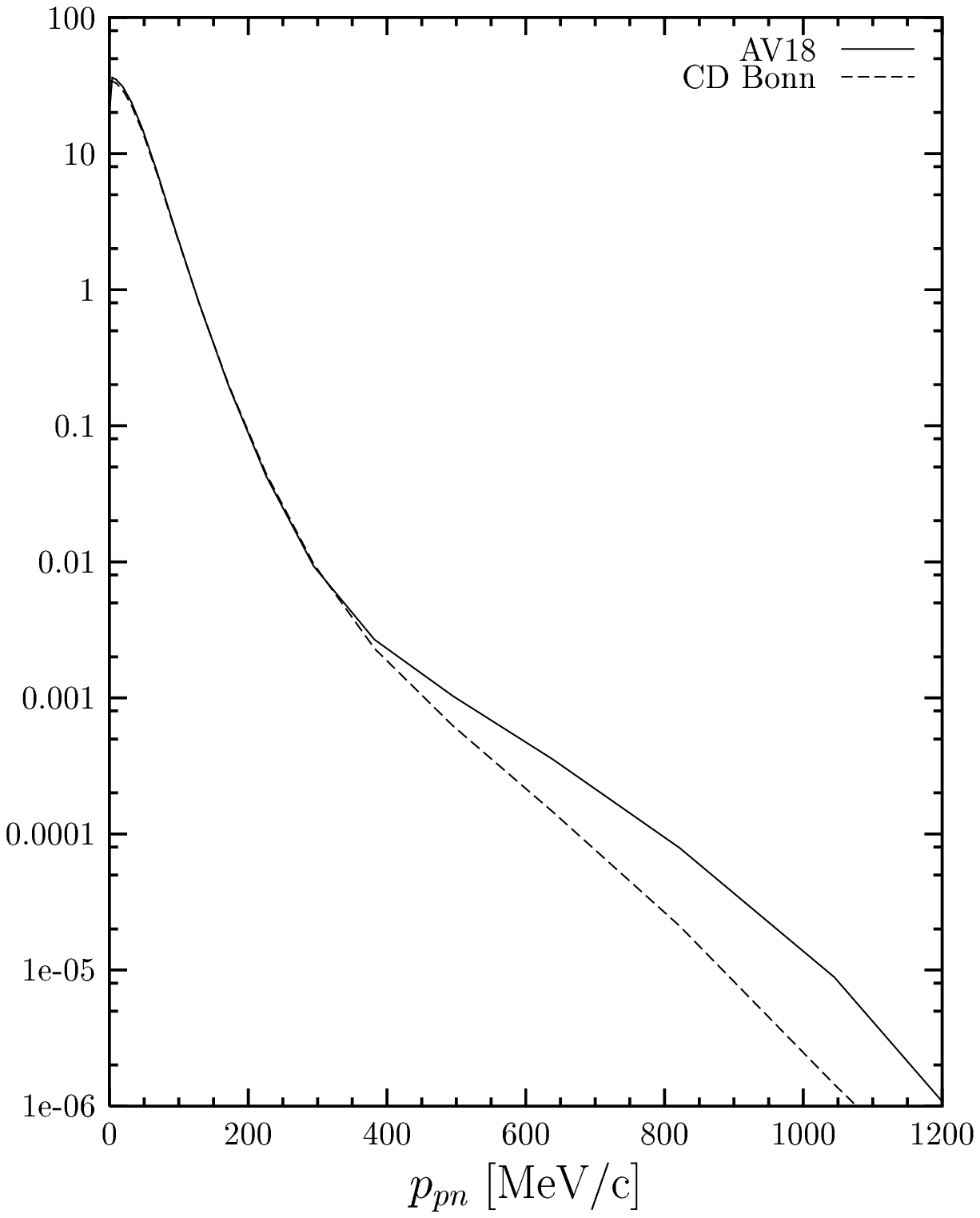,angle=0,scale=0.4}
\psfig{file=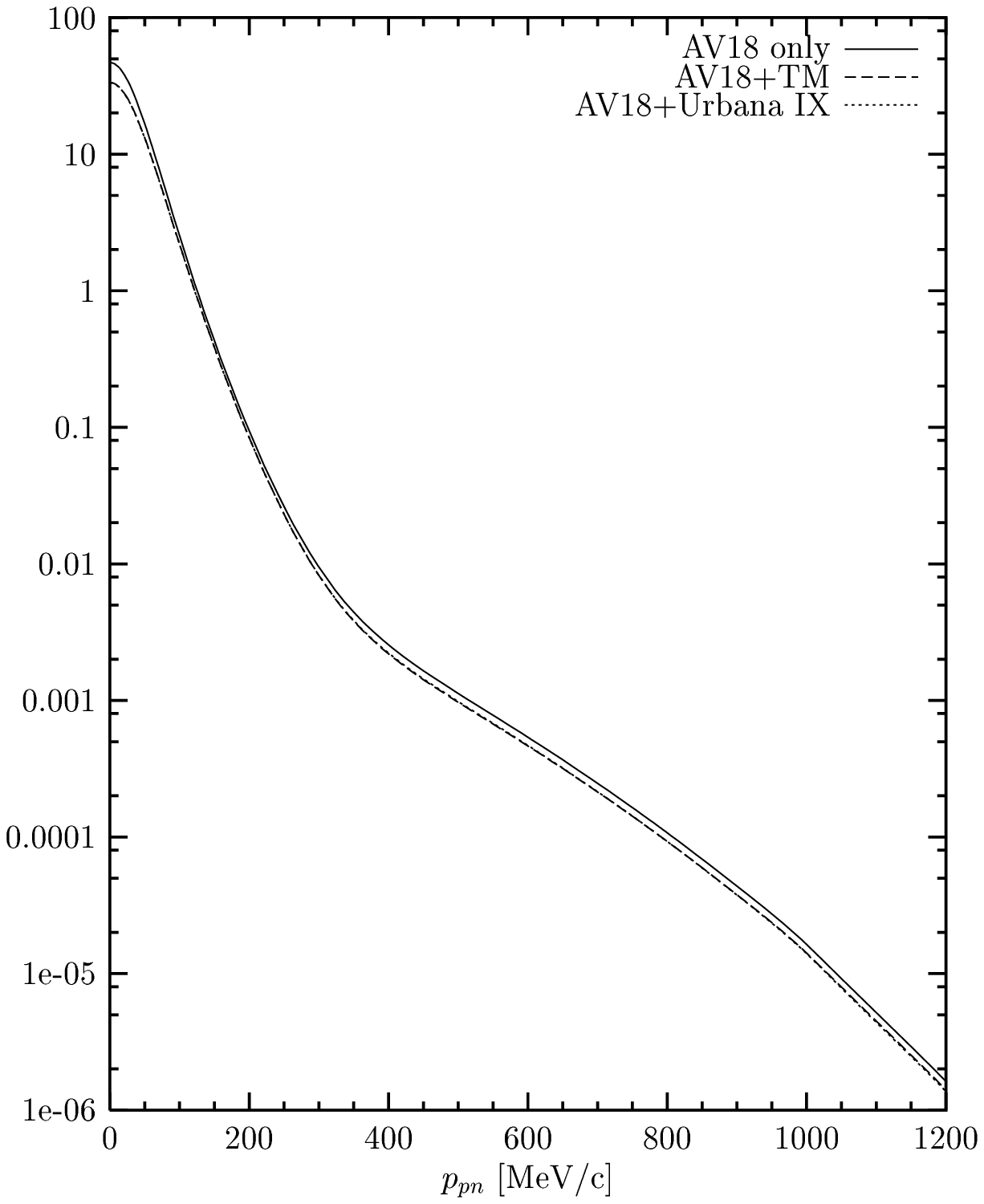,angle=0,scale=0.4}
\end{center}
\caption{
The expression in Eq.~\ref{eq:6} against the relative momentum $p$
for a $pp$ pair (left) and for a $np$ pair (right). Both refer to NN forces
only, whereas the figure in the middle includes in addition 3N forces.
 }
\label{fig:spe.npp}
\end{figure}

%%%%%%%%  FIGURE 12 
\begin{figure}
\begin{center}
\psfig{file=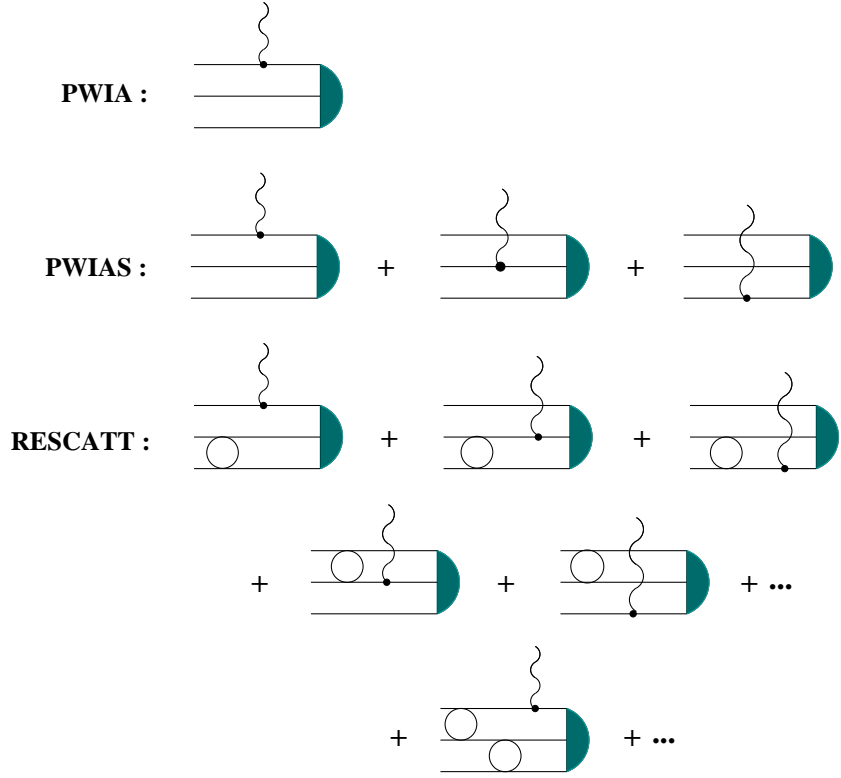,angle=0,scale=0.6}
\end{center}
\caption{
The matrix elements of Eq.~\ref{eq:12} contributing to PWIA, PWIAS
and to the infinite number of RESCATT processes. The first diagram
in the group RESCATT is the expression `` $t G_0 $ ''.
 }
\label{fig:diagram}
\end{figure}

%%%%%%%%  FIGURE 13 
\begin{figure}[htb]
\begin{center}
\psfig{file=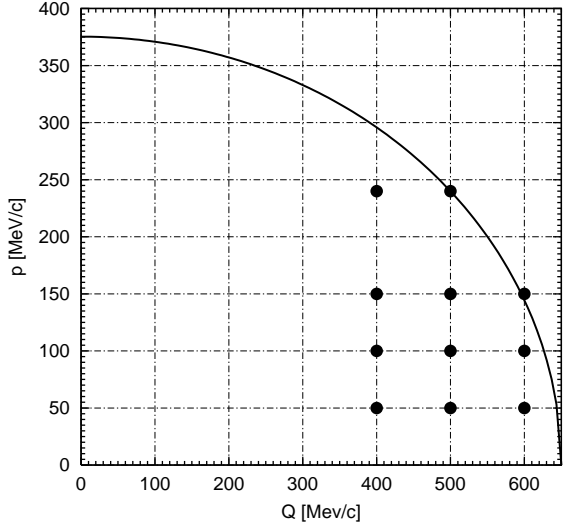,angle=-90,scale=0.5}
\end{center}
\caption{The restrictions for relative NN momenta $p$ as a function
of photon momenta $\vert \vec Q \vert $ choosing $E_{c.m.}^{3N}$= 140 MeV.}
\label{fig:ellipse}
\end{figure}
                                                                                
%%%%%%%%  FIGURE 14
\begin{figure}
\begin{center}
\psfig{file=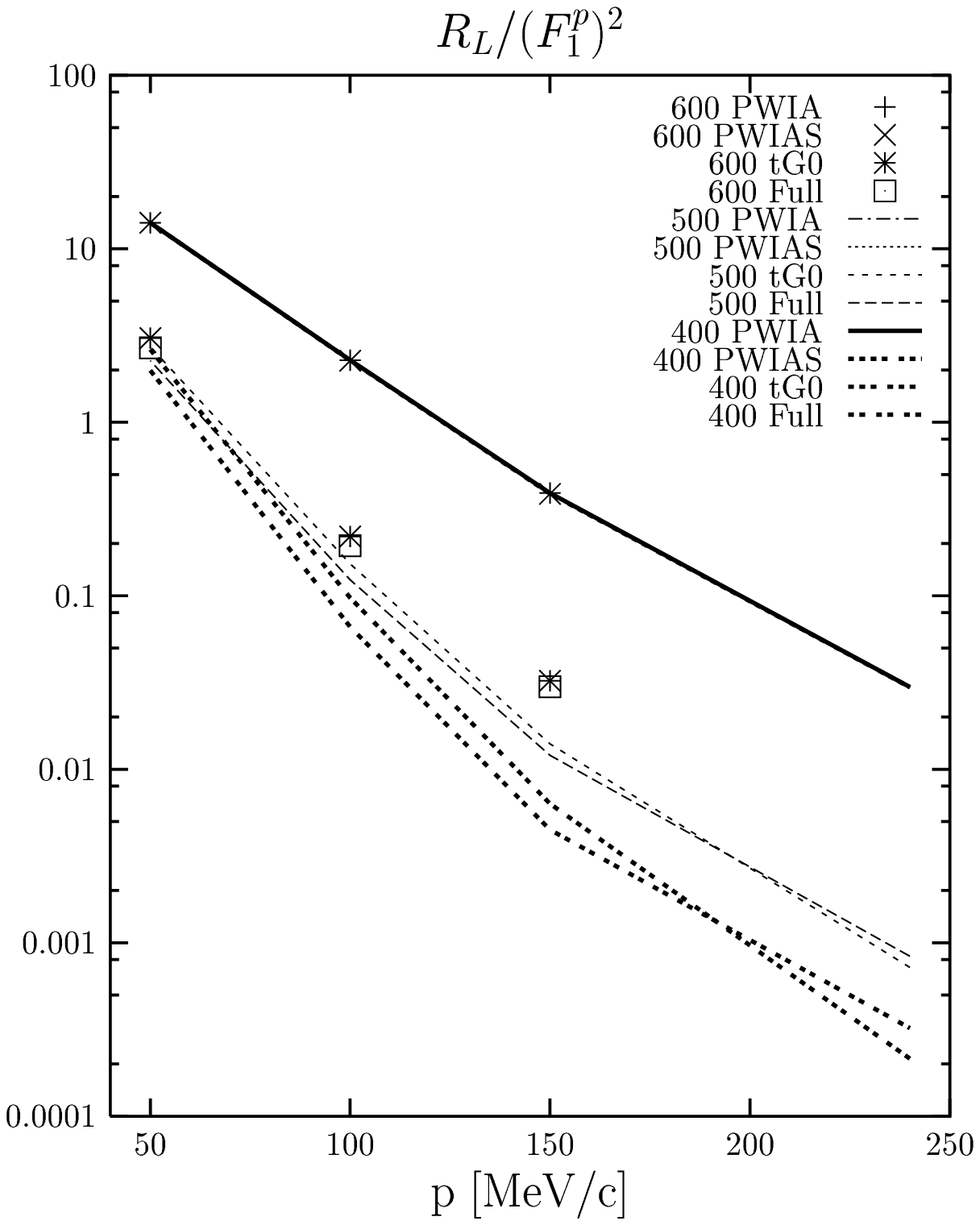,angle=0,scale=0.6}
\psfig{file=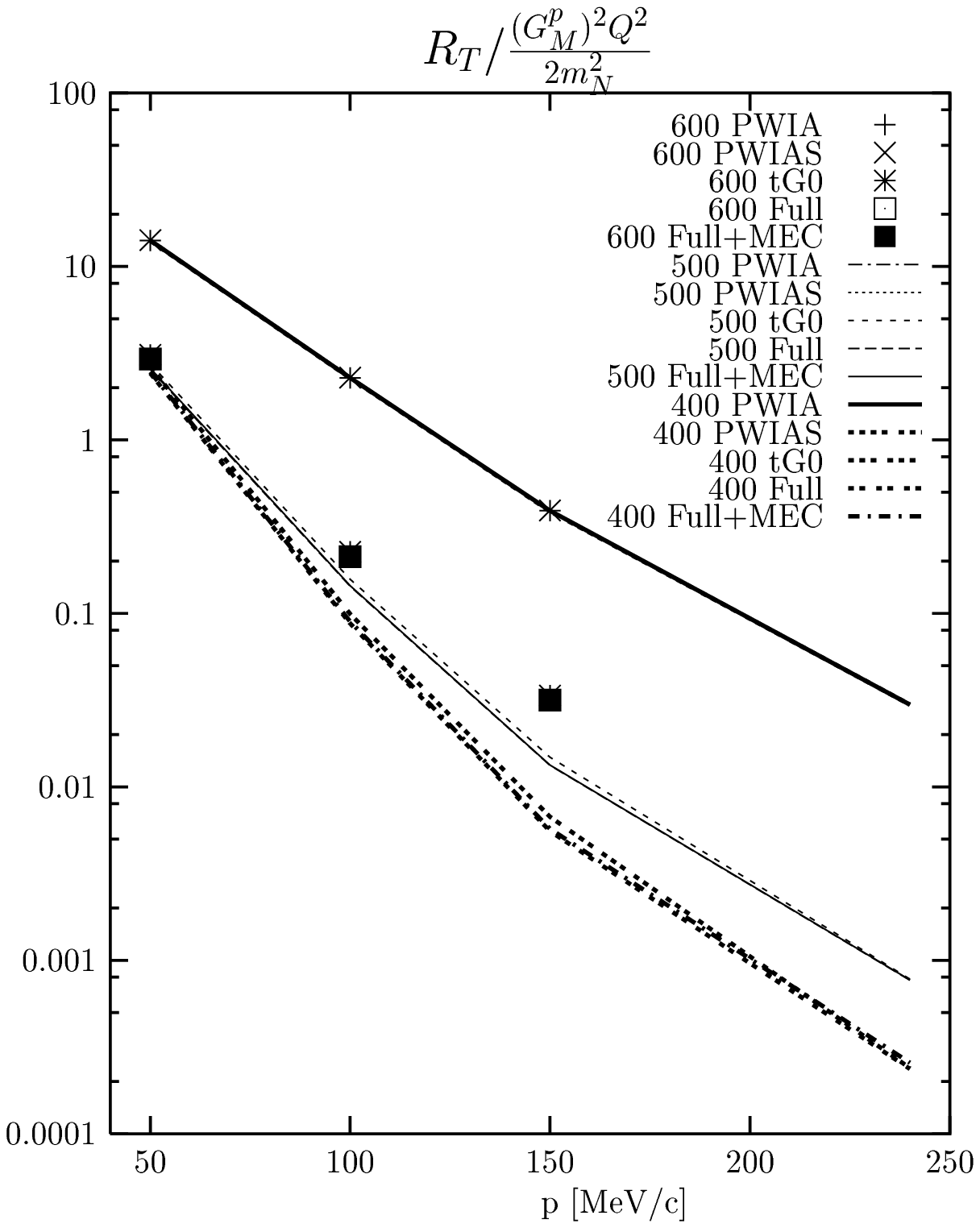,angle=0,scale=0.6}
\end{center}
\caption{$R_L/( F_1^p )^2 $ (up)
and $(2 m_N^2 R_T)/( G_M^p Q)^2 $ (down)
against the $np$ relative momentum $p$
for various treatments of the final 3N state and the $\vert \vec Q
\vert$-values 400, 500 and 600 MeV/c. MEC effects are negligible as shown.
 }
\label{fig:ppn.n0}
\end{figure}

%%%%%%%%  FIGURE 15               
\begin{figure}
\begin{center}
\psfig{file=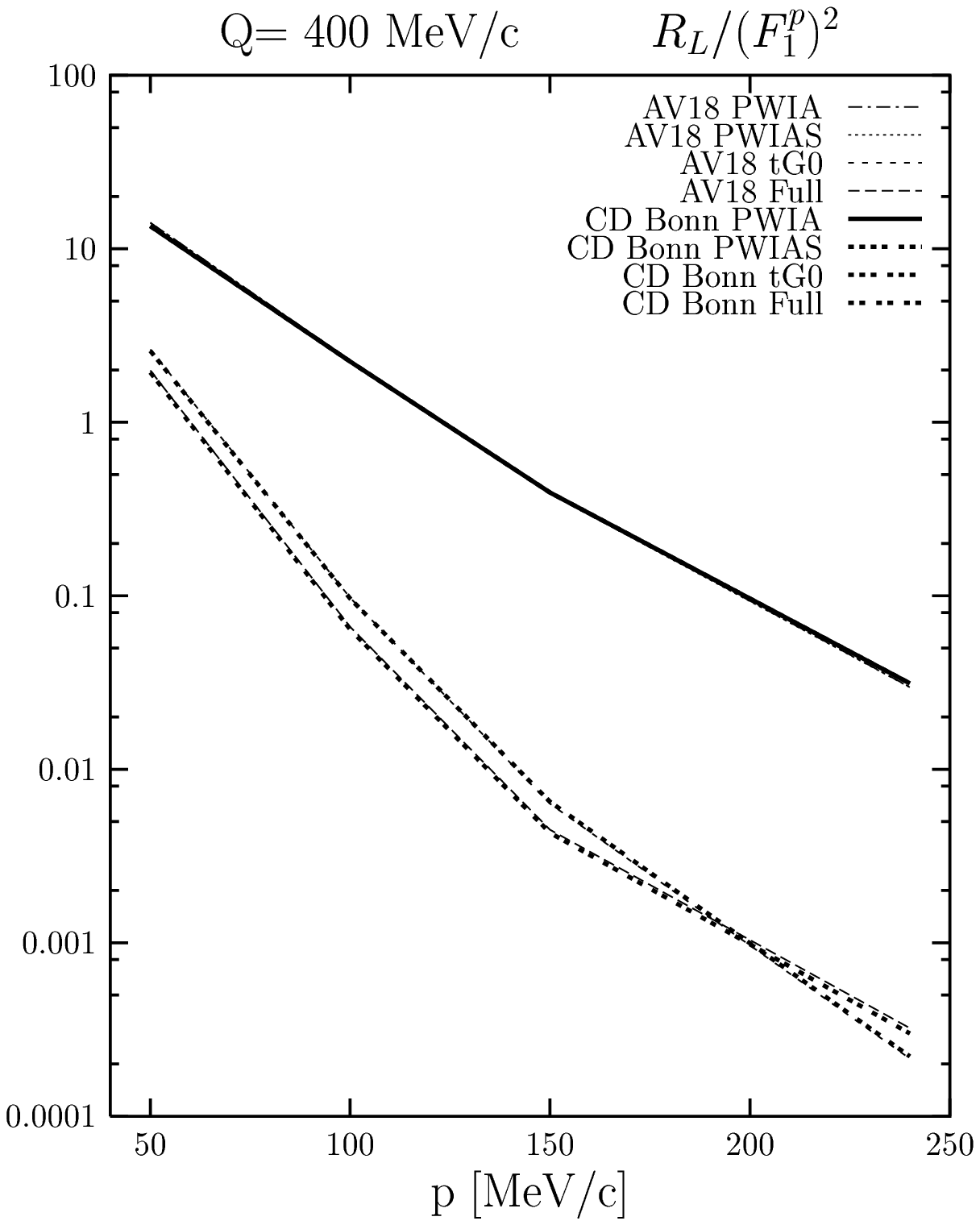,angle=0,scale=0.5}
\psfig{file=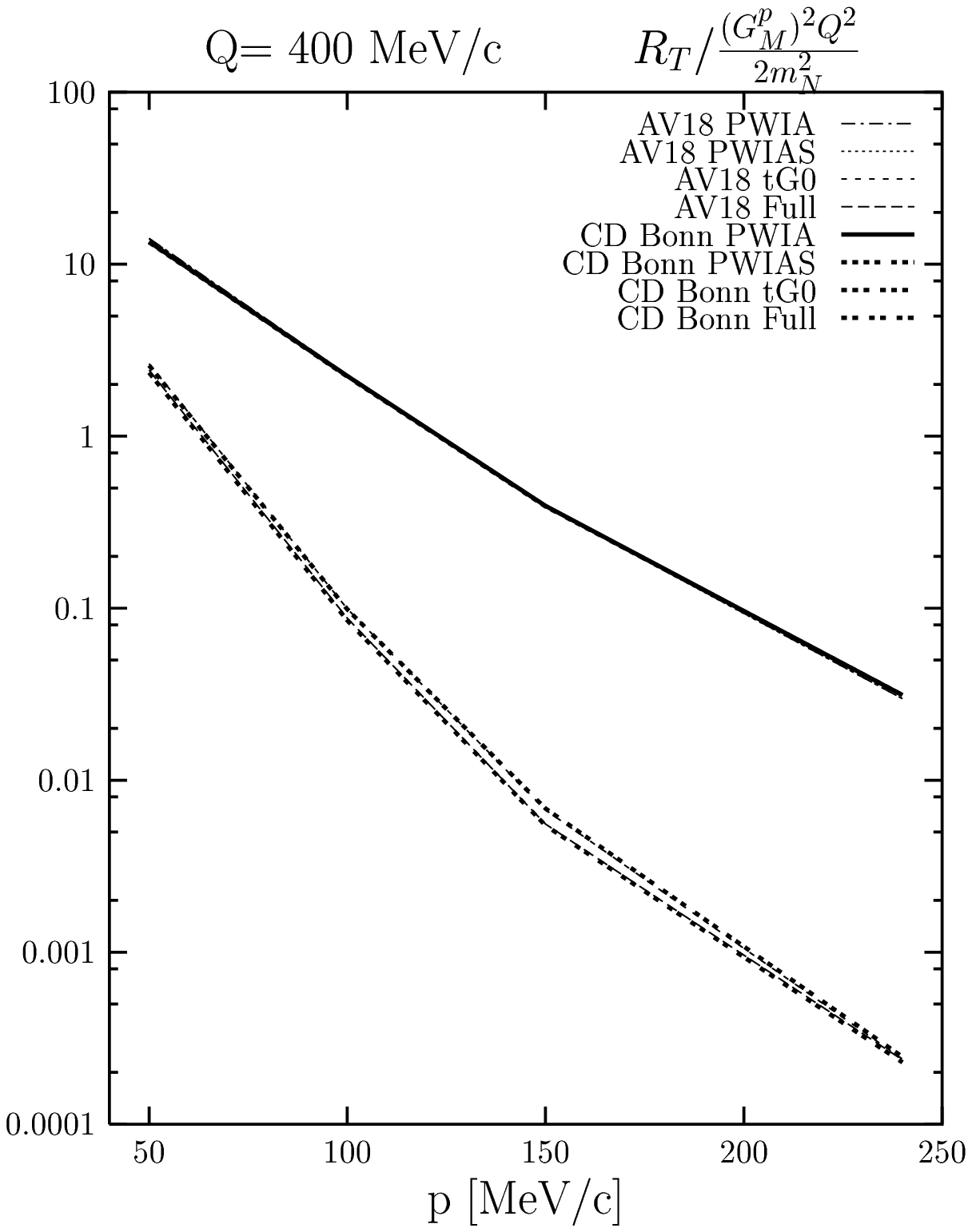,angle=0,scale=0.5}
\end{center}
\caption{
$R_L/( F_1^p )^2 $ (up)
and $(2 m_N^2 R_T)/( G_M^p Q)^2 $ (down)
against the $np$ relative momentum $p$
for various treatments of the final 3N state
and two different potentials at one $\vert \vec Q
\vert$-value.}
\label{fig:ppn.n0.2}
\end{figure}

The Argonne-Illinois-Los Alamos collaboration has explored with the help
 of the Greens-Function Monte Carlo method the low energy spectra of
 light nuclei up to A=8~\cite{ref10}. We see in Fig.~\ref{fig:spec2}
the pure AV18 predictions, which
 are rather far away from the data and the impressive shift of theory
 towards the data by adding the Urbana IX 3N force. But we still observe
deviations, which can be resolved by including additional terms in the
3NF \cite{ref10a}.
                                                                                Further important tests of the nuclear Hamiltonian without and with 3N forces
 are scattering processes. For three nucleons solutions for the
continuum are by far most developed. The Faddeev scheme~\cite{ref11} and the hyperspherical
harmonic method~\cite{ref12} provide very accurate solutions.
We illustrate the state of
 art with several cross sections and refer the reader to~\cite{ref11}
and more recent
 papers~\cite{ref13,ref14,ref15} for a larger overview and for the very many
spin observables,
 which probe our present day understanding of the dynamics in a very
 sensitive manner.
 Fig.~\ref{fig:ndtot} shows the $nd$ total cross section, which below about 100
MeV is nearly
 perfectly described even without 3N forces. Only at the higher energies
 small discrepancies appear, which are however significantly reduced
 including 3N forces. At very low energies the inclusion of the $pp$ Coulomb
 force is under control. For a survey on the beautiful agreement of the angular
 distribution in elastic $pd$ scattering with precise data we refer
to~\cite{ref12}.
 Thereby 3N force  effects are tiny. This remains true also at somewhat higher
 energies as shown in Fig.~\ref{fig:fig9n}, where the theory does not include
the Coulomb forces nor 3N forces. At 65 and 135 MeV theory based on NN forces
only clearly
 underestimates the data in the minima whereas the inclusion of 3N forces leads
 to  a very good agreement. This is shown in
Figs.~\ref{fig:e65p.elastic.ds} and \ref{fig:e135.elastic.ds}.
 Thereby the results, as shown by two bands, are very stable under
 exchange of NN and 3N force  combinations (which of course describe the
 $^3$H binding energy). Finally we show in Figs.~\ref{fig:fig35n}
and \ref{fig:fig39n} some break-up cross sections
 along the kinematical locus  as a function of  a suitably defined
arclength $S$ \cite{ref11}.
 In general the agreement is good and at  those low energies 3N force
 effects studied up to now are insignificant.
 
A prerequisite to a theoretical analysis of $^3{\rm He}(e,e'{\rm NN})$
 data
 is  the  understanding of  the d(p,NN) reaction in the full
 phase space. Therefore 4$\pi$-measurements of the latter process,
not only at certain selected regions in phase space,
 are quite
 important to test the  theory, before conclusions can be drawn from an
 analysis of the $^3{\rm He}(e,e'{\rm NN})$ reaction.
                                                              
However, overall one can already say now that the dynamical picture
with  high precision  NN forces
 and adjusted 3N forces works reasonably well (with room for
improvements) and provides a good basis to analyze electromagnetically
 induced processes.
 
\section{Two-Nucleon Correlation Functions}
 
Based on fully converged solutions of the Faddeev-Yakubovsky equations
 we present two-nucleon correlation functions for ground state wave functions
$\vert \Psi JM \rangle $  of the simplest type, namely averaged over
two-body partial wave states:
 
\begin{eqnarray}
C(r)= { 1 \over { 2 J+1}} \sum _M \langle \Psi JM \vert \delta ( \vec r
- \vec r _{ij} )   P_{ij} \vert \Psi JM \rangle
\label{eq:1}
\end{eqnarray} 

%%%%%%%%  FIGURE 16
\begin{figure}
\begin{center}
\psfig{file=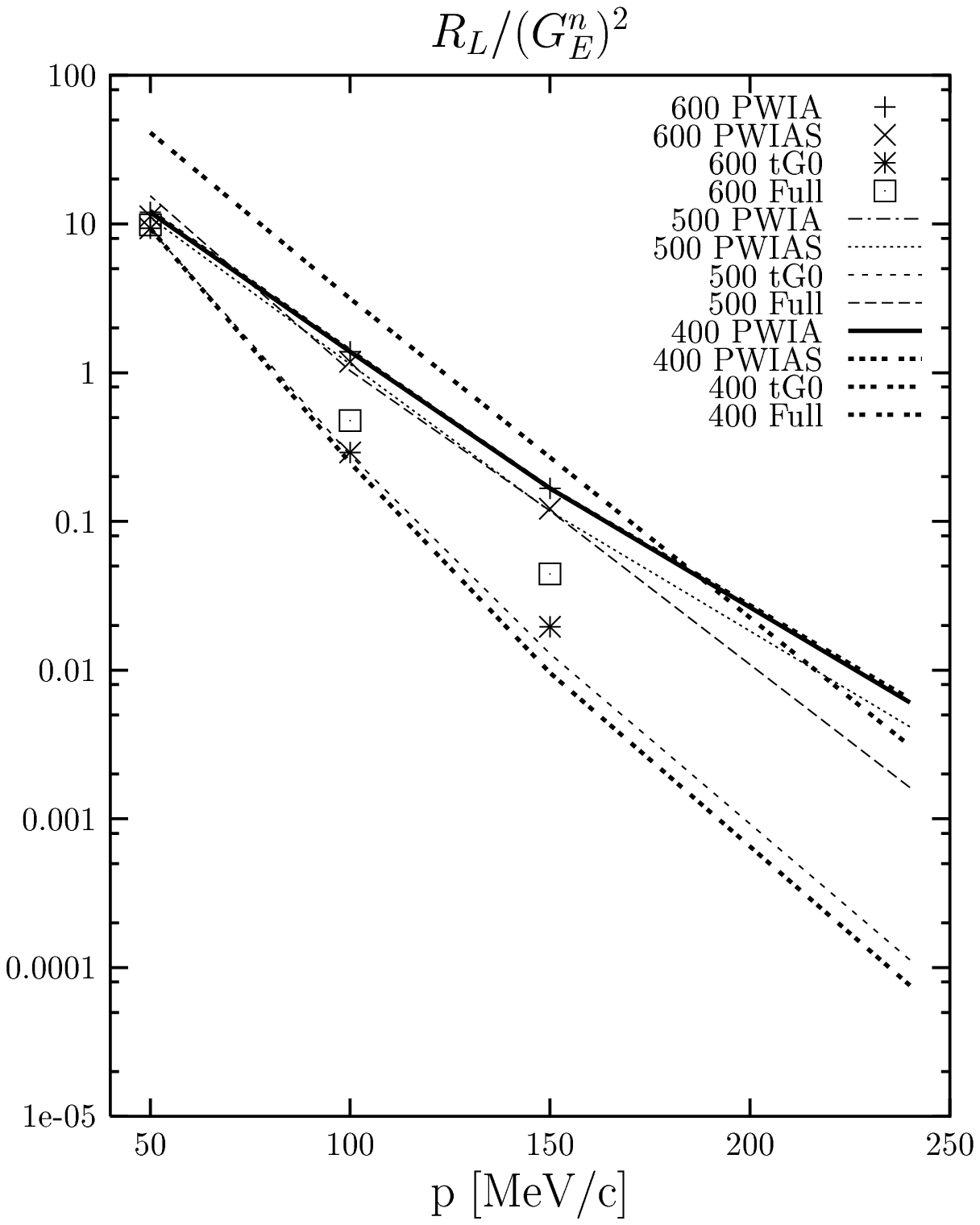,angle=0,scale=0.5}
\psfig{file=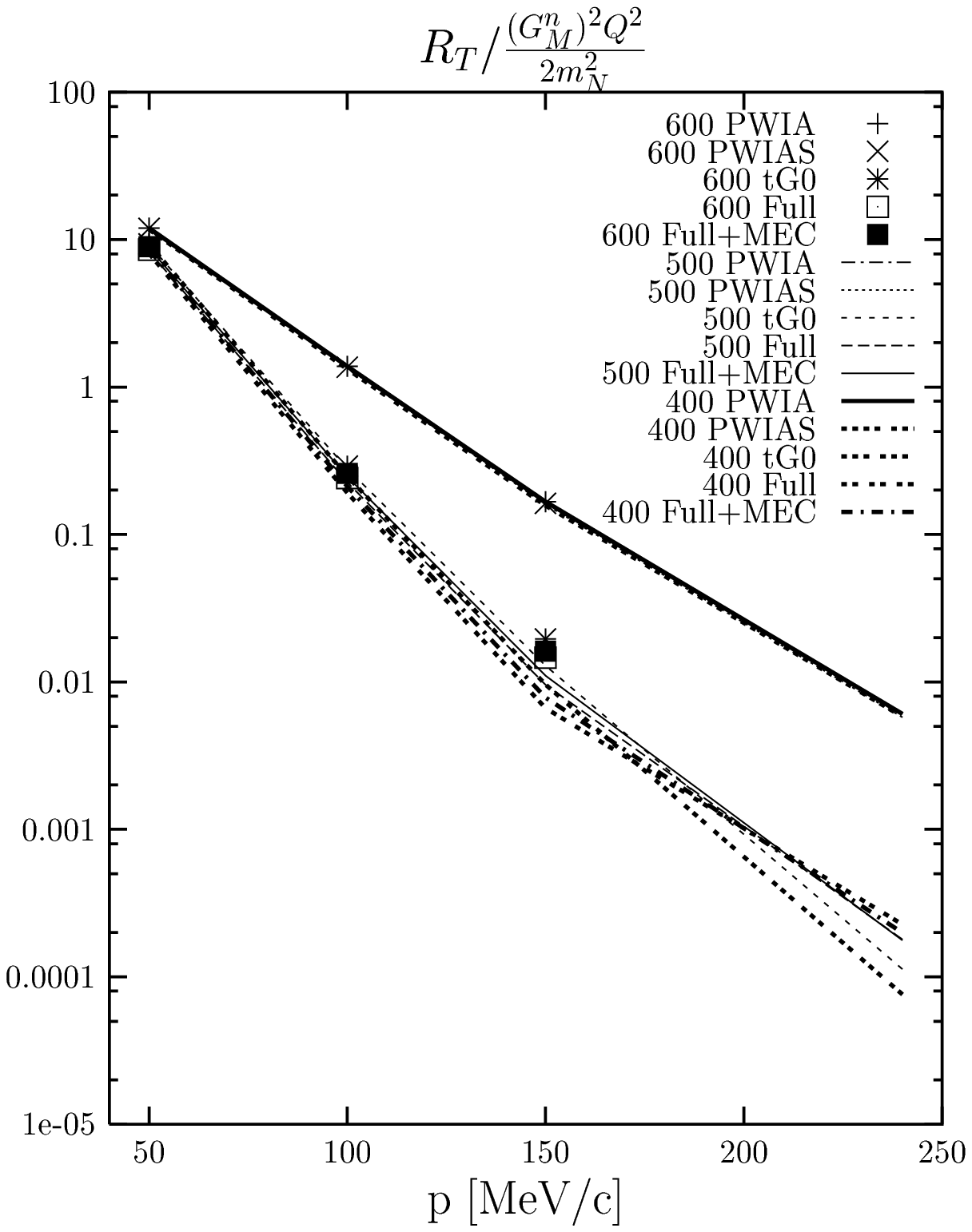,angle=0,scale=0.5}
\end{center}
\caption{$R_L/( G_E^n )^2 $ (up)
and $(2 m_N^2 R_T)/( G_M^n Q)^2 $ (down)
against the $pp$ relative momentum $p$
for various treatments of the final 3N state and some $\vert \vec Q
\vert$-values.  MEC effects are negligible as shown.
 }
\label{fig:npp.n0}
\end{figure}

%%%%%%%%  FIGURE 17 
\begin{figure}
\begin{center}
\psfig{file=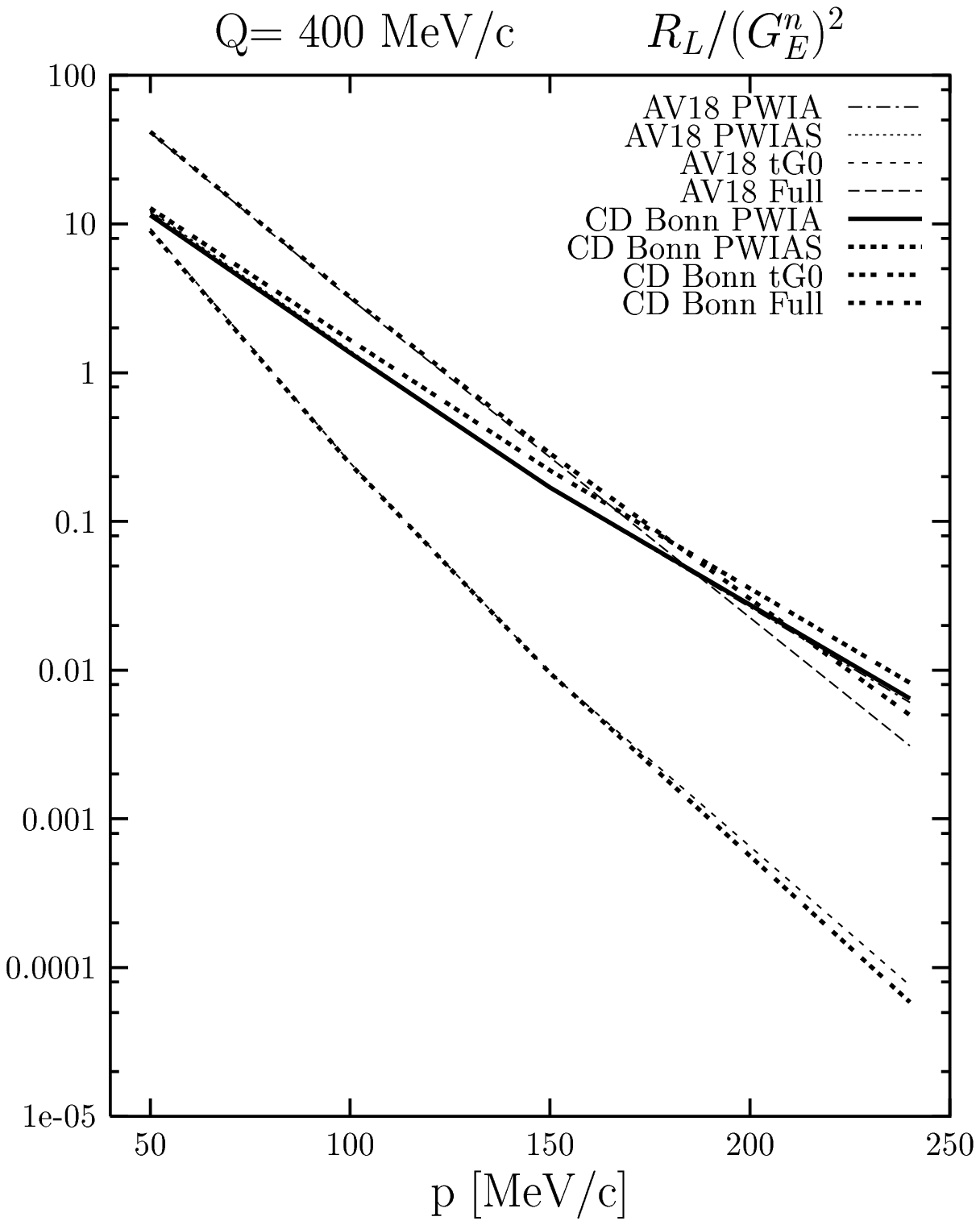,angle=0,scale=0.5}
\psfig{file=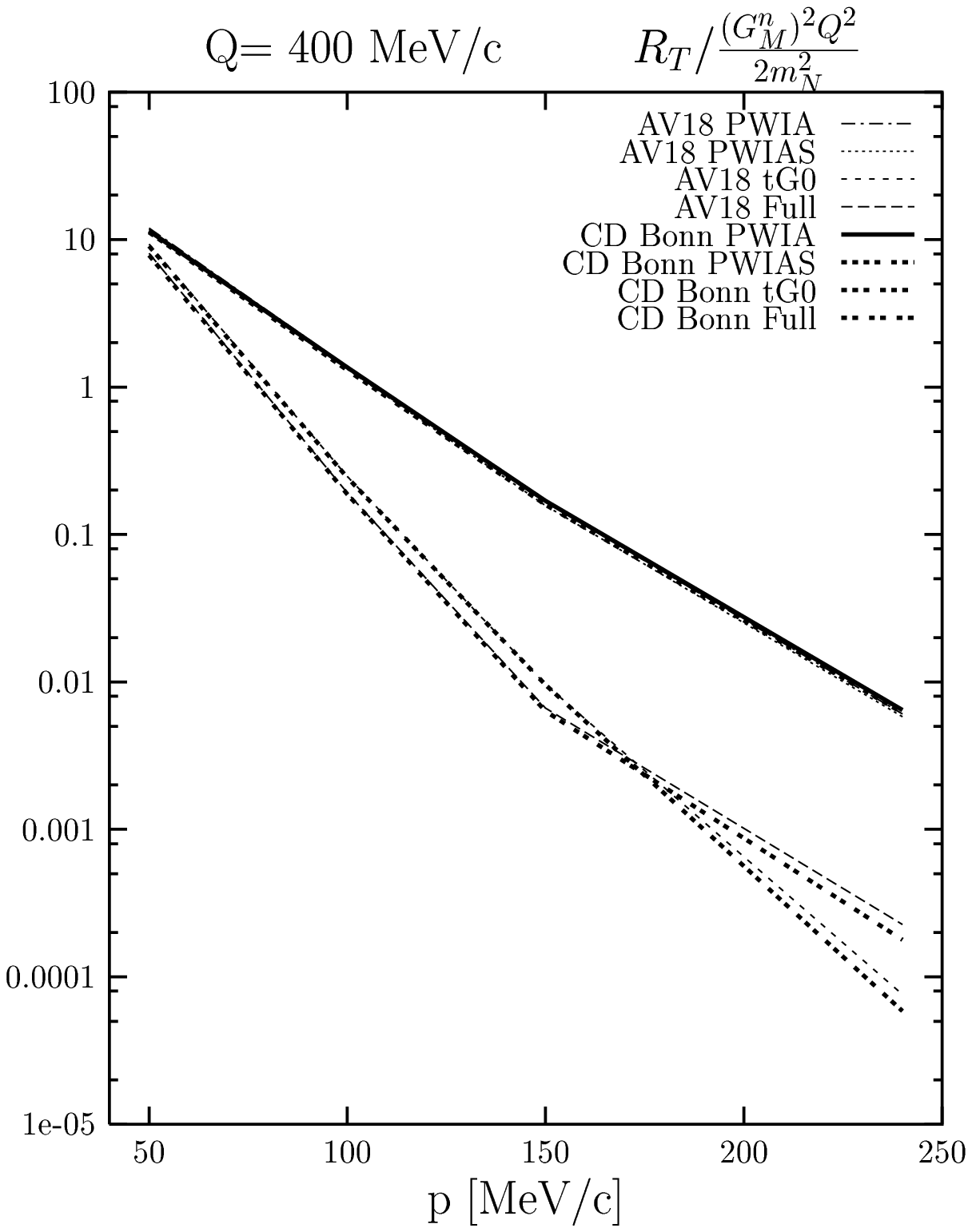,angle=0,scale=0.5}
\end{center}
\caption{
$R_L/( G_E^n )^2 $ (up)
and $(2 m_N^2 R_T)/( G_M^n Q)^2 $ (down)
against the $pp$ relative momentum $p$
for various treatments of the final 3N state
and two different potentials at one $\vert \vec Q
\vert$-value.}
\label{fig:npp.n0.2}
\end{figure}

%%%%%%%%  FIGURE 18   
\begin{figure}[htb]
  \begin{center}
    \parbox[b]{0.5\linewidth}{
      \includegraphics[width=0.8\linewidth]{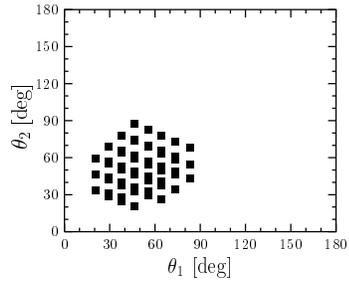}}
    \parbox[b]{7mm}{~}
    \parbox[b]{0.3\linewidth}{
      \caption[experimental setup]{\label{fig:ee}
      The $\theta_1$-$\theta_2$ regions in 3N
      phase space, where 3N force effects are larger than 40 \%.
      The relative azimuthal angle is $ \phi_{12} \le 50 ^\circ$.
      }
      }
  \end{center}
\end{figure}
 
%%%%%%%%  FIGURE 19 
\begin{figure}
\begin{center}
\psfig{file=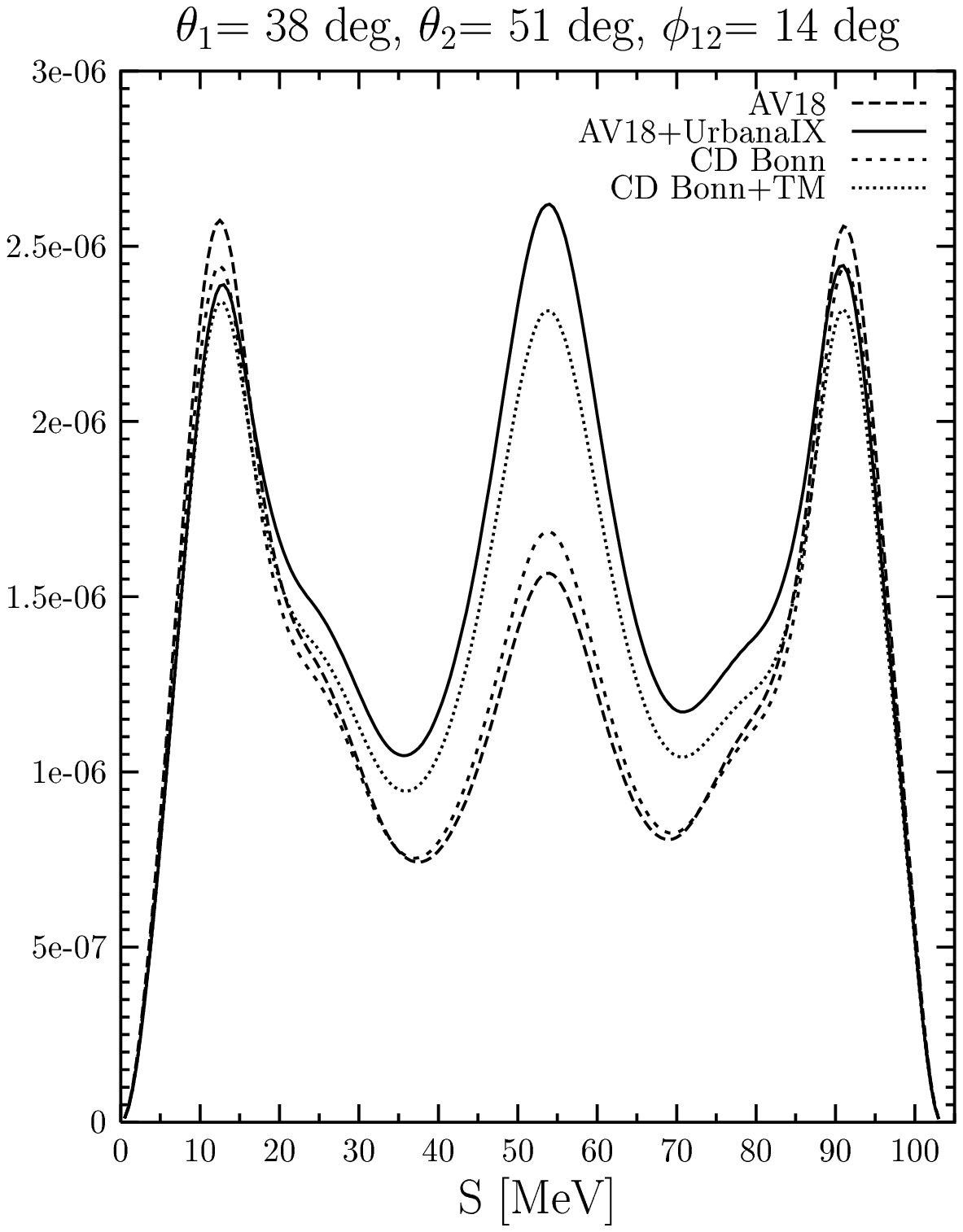,angle=0,scale=0.5}
\psfig{file=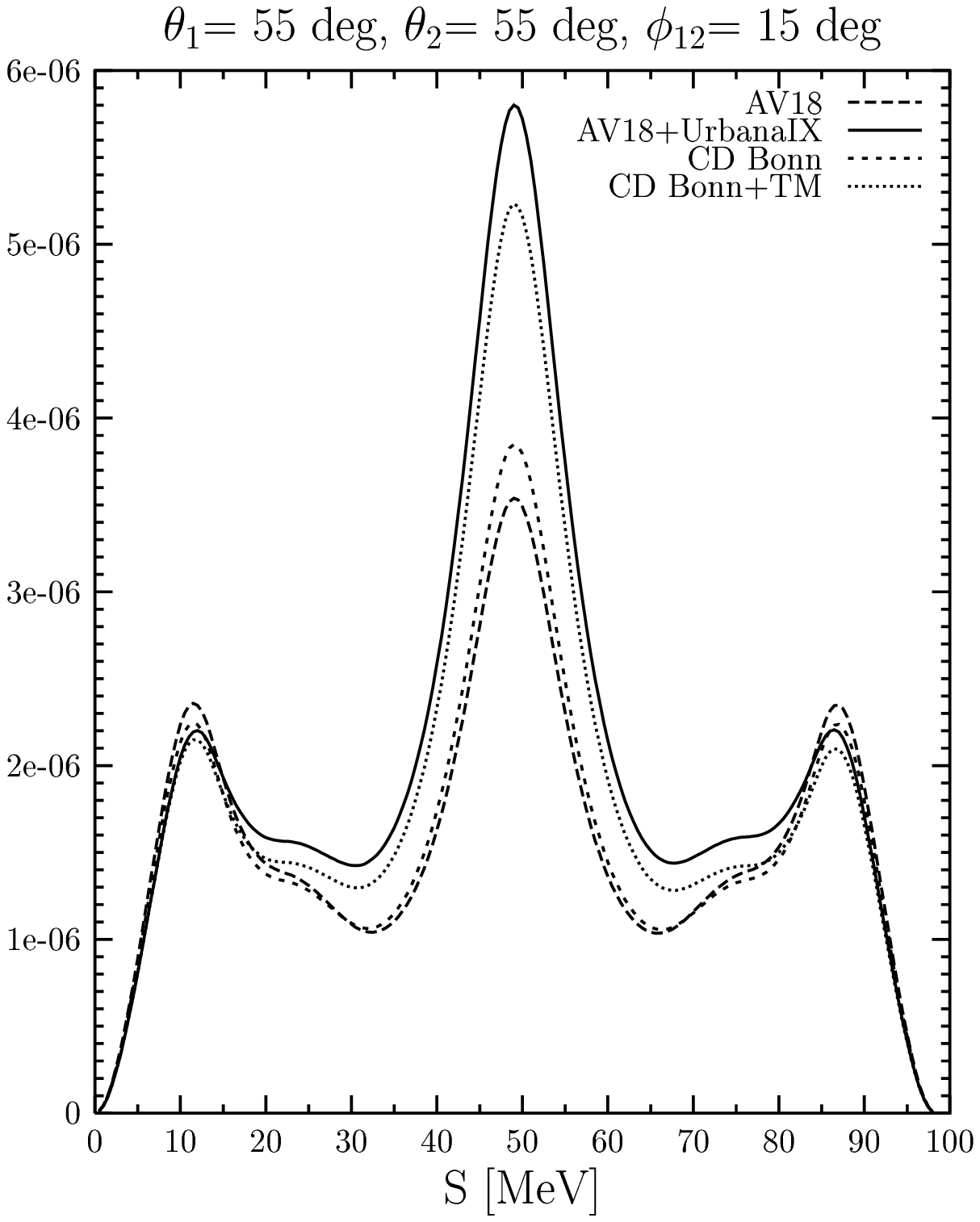,angle=0,scale=0.5}
\end{center}
\caption{Three-body differential photodisintegration cross sections
$ d^3 \sigma/(d\Omega_1 d\Omega_2 dS ) $ [${\rm fm}^2/({\rm sr}^2{\rm
MeV})$]
of $^3$He along the kinematical
locus for different combinations of angles.
}
\label{fig:c1-6}
\end{figure}
 
%%%%%%%%  FIGURE 20 
\begin{figure}
\begin{center}
\psfig{file=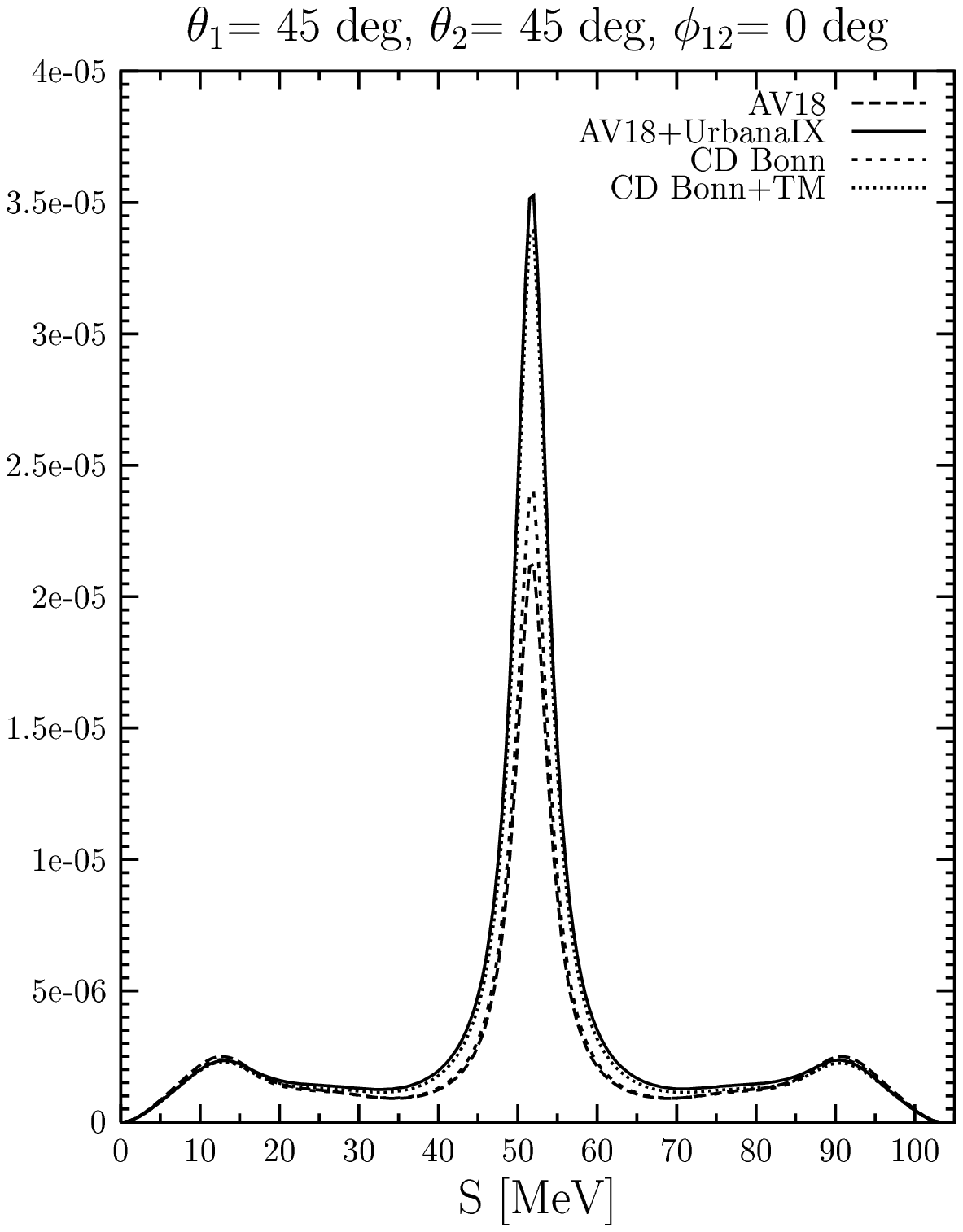,angle=0,scale=0.5}
\end{center}
\caption{The same as in Fig.~\ref{fig:c1-6} for a specific breakup
configuration, where two nucleons leave with equal momenta leading to a
FSI peak.
}
\label{fig:c1-6n}
\end{figure}
                 
\noindent
Here $\vec r_{ij}$ is the operator of a pair distance and  $P_{ij}$
the projector on a $pp$ or $np$ pair.
They are plotted (with arbitrary overall normalization)
in Fig.~\ref{fig:crnppp}\ for  $d$,
$^3$He and $^4$He
    and choosing two NN potentials  AV18 and CD Bonn. At short distances
 below about 1~fm the predictions of the two potential
are quite different. The curves
 are very much similar for the three nuclei, what suggests that this
 will essentially remain true also for heavier systems. They all peak
 at about 1 fm. Fig.~\ref{fig:cr3nf} compares directly predictions for the three nuclei,
 all normalized to each other in the peak value. We see a nearly perfect
 overlap except in the tails, where the difference in separation energies
 show up. Finally the addition of 3N forces has no visible influence up
 to the radii somewhat larger than 1~fm, where binding effects have to
appear.

 Since wave functions are no observables, in consequence
 the wave function property $C(r)$
 is not an  observable either. Wave functions enter into observable response
 functions as they occur for instance in electromagnetically induced
 processes, but they are accompanied by final state continuum wave functions,
 which are also correlated and, very importantly, they come together
 with current operators, which should be consistent to the nuclear forces.
 Consistent ingredients of the nuclear matrix elements should lead to the
 same observables (response functions) even if different NN force
 parameterization have been chosen. At present mostly AV18 NN forces
 and  consistent currents  are being used. Thus work remains to be
done to understand possible model dependences.
                                
\section{The Coulomb Sum Rule}

It has been known for a long time~\cite{ref23,ref24} that the Coulomb sum rule is an approach
to isolate the Fourier transform of the correlation function $C(r)$.
 Let's define the Coulomb sum as
\begin{eqnarray}
S_L \equiv \int _ {\omega _{min}} ^ \infty d \omega R_L ( \omega , \vert \vec  Q \vert )
\label{eq:2}
\end{eqnarray}
 
Then using the standard expression for the longitudinal response
function  $R_L$
in inclusive electron scattering together with the closure relation one
 easily finds \cite{ref25}
 
\begin{eqnarray}
S_L =  { 1 \over 2 } \ \sum _M   \langle \Psi JM \vert \rho ^\dagger \rho
 \vert \Psi JM \rangle
-  { 1 \over 2 } \  \sum _M \vert  \langle \Psi JM \vert \rho \vert \Psi JM \rangle \vert ^2
\label{eq:3}
\end{eqnarray}

This very nice intermediate result shows that the final state
 interactions have been totally removed and only ground state expectation
 values remain. Separating the density operator $\rho$ into single particle
 and two- and more-particle parts one easily arrives at
     
\begin{eqnarray}
S_L = Z G_E ^p (\vec Q ) ^2 + N  G_E ^ n ( \vec Q ) ^2 - Z^2 F_{ch} ^2 ( \vec Q
) + C( \vec Q ) + \tilde C ( \vec Q )
\label{eq:4}
\end{eqnarray}
where $G_E ^{p,n} $  are the $p,n$ electric form factors ( neglecting the time
 component of the four vector  dependence in $Q^2$), $F_{ch}$ is the elastic
 charge form factor of the target nucleus, $\tilde C  (\vec Q) $
 arises from the two- and more-particle densities and
 
\begin{eqnarray}
C (\vec Q) \equiv \int d^3 r e ^{ i \vec Q \cdot \vec r } C(r)
\label{eq:5}
\end{eqnarray}
is the quantity related to $C(r)$ suitably augmented by the nucleon
electromagnetic form factors \cite{ref25}.
 
%%%%%%%%  FIGURE 21 
\begin{figure}
\begin{center}
\psfig{file=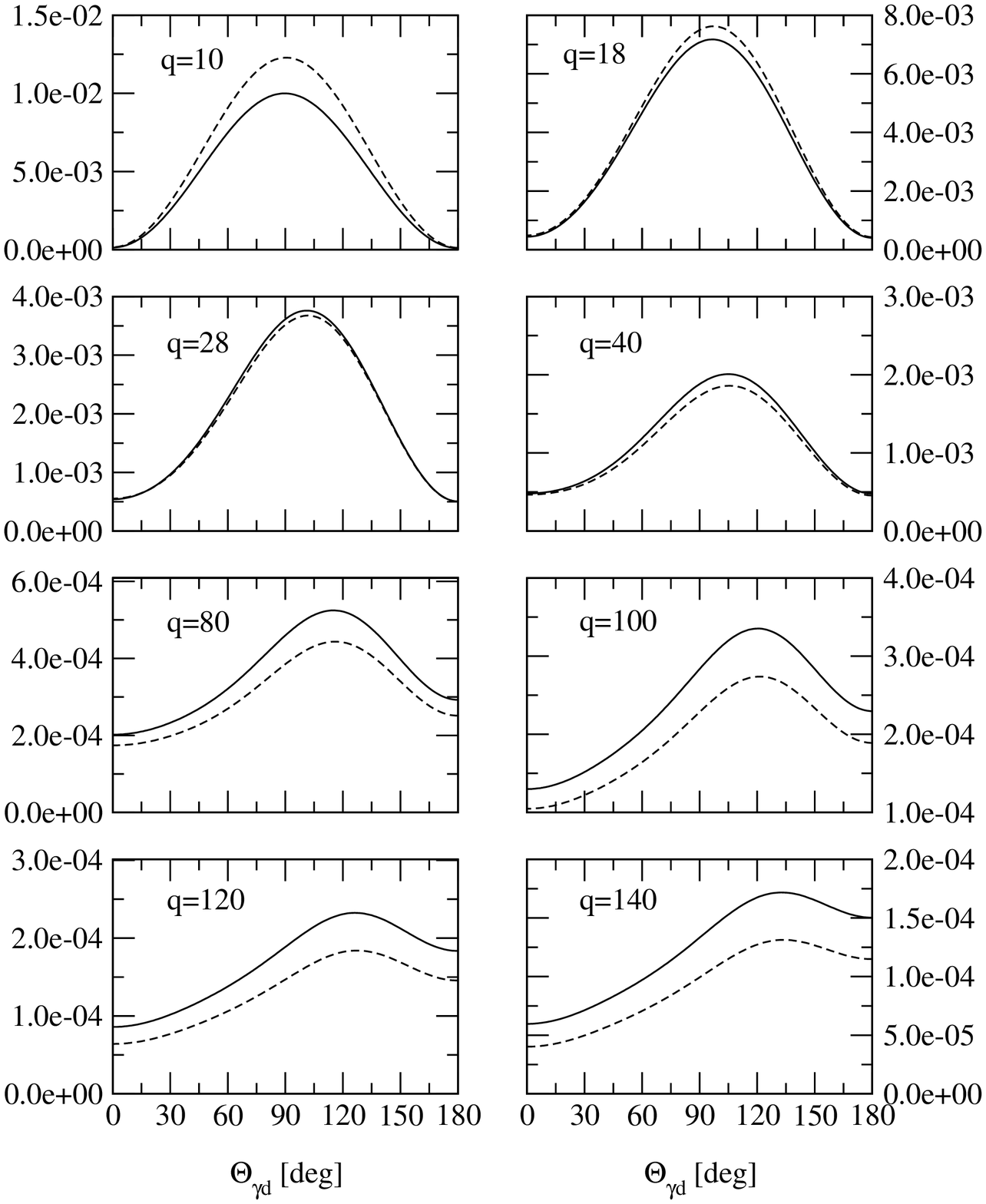,angle=0,scale=0.7}
\end{center}
\caption{The $pd$ photodisintegration breakup cross section
$ d \sigma/d\Omega_d $ [ ${\rm fm}^2/{\rm sr}$ ]  of $^3$He
based on AV18 alone (dashed curve) and AV18+Urbana~IX (solid curve)
for various photon energies $q$ in MeV.
}
\label{fig:romek.new}
\end{figure}

There are certain obstacles to be overcome. The integral in
Eq.~\ref{eq:2} requires
 an extrapolation to catch all of the integrand above the quasi elastic peak.
 Up to now the available data leave too much room for ambiguities in the
 extrapolation~\cite{ref25} and precise data at some more higher $\omega$-values would be
welcome. Also the first three terms on the right hand side of
Eq.~\ref{eq:4} cancel
 strongly~\cite{ref25} leading to enhanced error bars for $C$ and $\tilde C$.
Therefore high accuracy
 measurements of $R_L$ are needed. Above all this refers to the nuclei $^3$He
 and $^4$He, for which at present the most precise evaluation of the ground
 state wave functions are possible. On the theoretical side as has been shown
in~\cite{ref26} the relativistic corrections in the density operator
and two-body pieces therein play an important
role and cannot be considered as a small perturbation. This is an
interesting challenge for theory and experiment and deserves a renewed
 effort despite the intensive work in the past.
                              
\section{Exclusive Electron Scattering on $^3$He}

Another approach towards correlations is via electron induced
two-nucleon emission on nuclei, here on $^3$He. The most ideal
situation would be that one nucleon absorbs the photon and receives its
full momentum and all three nucleons leave the nucleus without any final
state interaction (FSI).
Then the measurement of the momenta of the two spectator
nucleons would display directly their momentum  distribution in $^3$He.
For this special kinematics the total spectator pair momentum has to be
zero and those two nucleons leave back to back and show directly the relative momentum
dependence within a pair of nucleons in the target nucleus $^3$He. Let us
number the nucleons such that the knocked out nucleon is number 1. Then
one would probe directly the expression related to the $^3$He wave function
 
\begin{eqnarray}
\sum_M \sum_{ m_1 m_2 m_3 } \vert \Psi (\vec p , \vec q =0 ) \vert ^2
\label{eq:6}
\end{eqnarray}
where $\vec p = {1 \over 2 } ( \vec k_2 - \vec k_3 ) $
and $\vec q = { 2 \over 3 } ( \vec k_1 -  { 1 \over 2 } (\vec k _2 + \vec k_3 ))$.
The $\vec k_i, i=1,2,3$ are the individual momenta.
This quantity is plotted in Fig.~\ref{fig:spe.npp}
for the two
spectator pairs $pp$ and $np$ and the two NN forces AV18 and CD-Bonn.
We see model independence below about $p=1$ fm$^{-1}$  and significant model
dependence at higher  $p$-values. 3N force effects are,  like for
C(r),  quite insignificant as shown in Fig.~\ref{fig:spe.npp} for a $np$
pair.

Unfortunately reality is far from that ideal situation. FSI
interferes very strongly. The eightfold differential cross
section has the well known form    

\begin{eqnarray}
{ {d ^8  \sigma}  \over { d \hat k ' d k_0 ' d\Omega _1 d \Omega _2 d S }}
=\sigma _{Mott} \,  ( v_L R_L + v_T R_T + v_{TT} R_{TT} + v_{TL} R_{TL}
) \, \rho
\label{eq:7}
\end{eqnarray}
where the  v's and $\rho$ are kinematical quantities, and the response
functions $R$ are
\begin{eqnarray}
R_L &=& \vert N_0 \vert ^2  \cr
R_T &=& \vert N_1 \vert ^2 + \vert N_{-1} \vert ^2 \cr
R_{TT} &=& 2 Re ( N_1 N_{-1} ^* ) \cr
R_{TL} &=& - 2 Re ( N_0 ( N_1 + N_{-1})^* )
\label{eq:8}
\end{eqnarray}
Here enter the nuclear matrix elements $N_i$, which are the spherical
components of
\begin{eqnarray}
N^ \mu = 3 \langle \Psi ^ {(-)} \vert j ^ \mu ( \vec Q ) \vert \Psi _{ ^3 He} \rangle
\label{eq:9}
\end{eqnarray}
with
\begin{eqnarray}
j^ \mu  ( \vec Q ) = j ^ \mu ( \vec Q ,1 ) + j ^ \mu ( \vec Q , 23 ) + \dots
\label{eq:10}
 \end{eqnarray}
Note that both 3N states in Eq.~\ref{eq:9} are fully antisymmetrized. Using a Faddeev
decomposition, $N^\mu$  can be written as
\begin{eqnarray}
N^\mu = N^\mu _{PWIAS} + N ^ \mu _{RESCATT}
\label{eq:11}
\end{eqnarray}
where                
\begin{eqnarray}
\label{nmatr}
 N^\mu _{PWIAS} &=& 3 \langle  \Phi ^ 0 _{\vec p \vec q } \vert (1 + P )
j^ \mu (\vec Q ) \vert \Psi _{ ^3 He} \rangle
\cr
N ^ \mu _{RESCATT}  &=&  3 \langle  \Phi ^ 0 _{\vec p \vec q } \vert (1 + P )
\vert U ^ \mu \rangle
\label{eq:12}
\end{eqnarray}
and $\vert U^ \mu  \rangle $  obeys a Faddeev-like integral
equation \cite{iiii}
\begin{eqnarray}
\vert U ^\mu \rangle = t G_0 ( 1 + P) j ^ \mu (\vec Q ) \vert \Psi _{ ^3 He} \rangle
+ t P G_0 \vert U ^ \mu \rangle
\label{eq:13}
\end{eqnarray}
The bra-vectors of Eq.~\ref{nmatr} is given by the free momentum eigenstates
\begin{eqnarray}
\vert \Phi ^ 0 _{ \vec p \vec q} \rangle
 = ( 1 - P _{23} ) \vert \vec p \rangle
\vert \vec q \rangle
\label{eq:14}
 \end{eqnarray}
antisymmetrized in the pair $(23)$,
$t$ is the NN t-operator and $G_0$ the free 3N propagator. Finally
P is the sum of a cyclic and anticyclic permutation of 3 objects. It is
instructive to display the physical content of Eq.~\ref{eq:12}
graphically in Fig.~\ref{fig:diagram}.
There the different treatments of the final state, PWIA, PWIAS,
``$tG_0$'' and ``full'' are explained.   
It is easy to show that in PWIA one finds
\begin{eqnarray}
R_L &=& G_E ^2 (\vec Q)  { 1 \over 2} \sum  \vert \Psi ( \vec p , \vec q=0 )
 \vert ^2
\cr
R_T &=&{  {\vec Q}^2 \over {2 m  _N ^2 } } G_M ^2 (\vec Q ) { 1 \over 2 }
\sum  \vert \Psi ( \vec p , \vec q=0 ) \vert ^2
\label{eq:15}
\end{eqnarray}
and  $R_{TT} = R_{TL} =0 $.
Thus $R_L$  and $R_T$  provide access to the same quantity $\sum \vert \Psi
 \vert ^2 $          up to
known factors. The most simple approximate FSI is
the action of $t$ within the spectator pair of nucleons, called `` $tG_0$''
below in Figs.~\ref{fig:ppn.n0}--\ref{fig:npp.n0.2}
Since we are restricting ourselves to a nonrelativistic
framework the total 3N c.m. energy should be below the pion threshold
at 140 MeV. Then for
given $\vert \vec Q \vert $  the p-values are restricted as shown
in Fig.~\ref{fig:ellipse}.  For three $\vert \vec Q \vert $-values,
400, 500, 600 MeV/c,
we compare in Fig.~\ref{fig:ppn.n0}
the quantity $ R_L/(F_1^p)^2 $ for
PWIA, PWIAS, `` $tG_0$'', and  using the complete FSI (called Full),
as a function of  the relative momentum of an outgoing $np$
pair. In PWIA $\vec p$ is related to the spectator pair (the
above mentioned pair 23). Consequently it is assumed that the
knocked-out nucleon is a proton.
Of course for PWIA there is no $\vert\vec Q \vert$-dependence and we see
directly the expression given in Eq.~\ref{eq:6}.
FSI has a tremendous effect for all three $\vert\vec Q \vert$-values,
however the simple approximation `` $t G_0 $ '' improves going to the
higher $\vert\vec Q \vert$-values. Even for higher $| \vec Q |$ values
FSI has a tremendous effect on $R_L$ (a reduction factor 10-100). At
least for the higher $| \vec Q | $ values the $t G_0$ approximation
might be reasonable, but still one can see effects of the neglected
FSI of spectator nucleons and hit proton.
In case of $R_T$  we show in Fig.~\ref{fig:ppn.n0} the ratio $ ( 2
m_N^2 R_T )/(G_M^p Q )^2 $.
In addition some curves include $\pi$ and $\rho$
exchange contributions consistent to AV18. Their effect is totally
negligible. Also in this case the approximation `` $tG_0$'' is a quite
good representation of the full FSI. Since that FSI correction of
first order in $t$ is very well under control given the fact that the
on-shell $t$ is fitted to the NN data, an analysis of future data
appears to be a rather reliable approach towards the quantity given in
Eq.~\ref{eq:6}.           
There is nice stability under exchange of the NN forces AV18 against CD
Bonn, as shown in Fig.~\ref{fig:ppn.n0.2}
for the example $\vert\vec Q \vert$= 400 MeV/c.
 
The situation is quite different if the hit nucleon is a neutron.
We show in Fig.~\ref{fig:npp.n0} the quantity $ R_L/(G_E^n)^2$.
In contrast to the case before, the approximate FSI treatment `` $t G_0 $ ''
is now totally different from the complete FSI result and thus
would be totally misleading. This is of course due to the smallness of
$G_E^n$. For $R_T$, however, the `` $t G_0 $ '' approximation
and `` Full '' are not far from each other and moreover the $\vert \vec Q \vert
$-dependence is relatively weak. Nevertheless for both $R$'s
the predictions are nearly independent
from the specific choice of the NN force as shown in
Fig.~\ref{fig:npp.n0.2}
and therefore the model dependence is weak.
 
All the curves in Figs.~\ref{fig:ppn.n0} and \ref{fig:npp.n0} refer to AV18
and the angle between $\vec p$ and $\vec Q$  has arbitrarily  been fixed
at $ 90 ^\circ$.
At the other angles the relation between the ``full''
result and the `` $t G_0 $ '' approximation changes, but remains
within the same order of magnitude.
 
Based on these results one has to state that there is no way to
access directly in that low energy regime $\sum \vert \Psi \vert ^2 $.  Nevertheless precise measurements would be extremely
informative to test the whole dynamical picture, forces and currents.
Choosing other kinematical conditions even in PWIA the $^3$He bound
state $\Psi (\vec p , \vec q )$
   is probed in such
a manner that both $\vec p$ and $\vec q$  vary.
Under the prerequisite that the $pd$ break-up process has been tested
carefully against theory the $^3{\rm He}(e,e'{\rm NN})$ reaction
for general kinematics
is  a perfect tool to probe the remaining unknown ingredient, the
current operator \cite{ref26a}.

\section{3N Force Effects in Photo-Induced Disintegration of $^3$He}

3N forces are required for a correct description of
binding energies. Will they also  play a role in electromagnetically induced
processes ? We already started a first investigation for $pd$ capture
processes in~\cite{golak2000} and would like to show new results for
photodisintegration of $^3$He, specifically at higher energies than the one
considered in~\cite{golak2000}. To that aim the Faddeev-like integral
equation in~\cite{iiii}
has to be modified. The nuclear matrix element for $^3$He $(\gamma,NN)$
\begin{eqnarray}
N^\mu = \langle \Psi ^ {(-)} _{ \vec p \vec q } \vert   j^\mu
\vert \Psi _{^3 He} \rangle
\label{eq:16}
\end{eqnarray}
can be written as
\begin{eqnarray}
N^ \mu = { 1 \over 2 } \langle \Phi ^ 0  _{ \vec p \vec q }
\vert ( 1 + t G_0 ) P \vert \tilde U ^ \mu \rangle
\label{eq:17}
\end{eqnarray}
where $ \vert  \tilde U ^ \mu \rangle $ obeys the Faddeev-like integral equation\begin{eqnarray}
\vert \tilde U ^ \mu \rangle = ( 1 + P )  j^\mu  \vert  \Psi _{^3 He} \rangle
+ ( t G_0 P + { 1 \over 2 } ( 1 + P ) V_{4} ^{(1)} G_0 ( 1 + t G_0 ) P) \vert
  \tilde U ^ \mu \rangle
\label{eq:18}
\end{eqnarray}
We use the Siegert theorem as described in~\cite{golak2000}. This includes some
of the
exchange currents. The operator $V^{(1)} _4 $ is that part of the 3N force,
which is symmetric under the exchange of particle 2 and 3.
 We scanned the whole phase space comparing the
full break-up cross  section evaluated with and without 3N force. As an
example we display in Fig.~\ref{fig:ee}  the $\theta_1, \theta _2 $
      regions where 3N force  effects are
larger than 40~\%. The relative azimuthal angle is $ \le 50 ^\circ $.
                                                                
As an illustration we show out of that phase space region
three more or less
arbitrarily selected break-up cross  sections along the $S$--curve
in Figs.~\ref{fig:c1-6}--\ref{fig:c1-6n}.
The peak in the middle is caused by small relative momenta in one pair
(FSI peak). In the other peaks the $^3$He wave function is probed at small
momenta. Measurements should validate or invalidate these sort of
predictions. Also in the $pd$ break-up process 3NF effects are clearly
visible as shown in Fig.~\ref{fig:romek.new} for various photon energies.

\section{Outlook}

One main theoretical challenge is to establish an electromagnetic
current operator, which is consistent to nuclear forces. Only then response
functions for electromagnetically induced processes can be put on
a firm ground. Precise experimental data on two-nucleon emissions induced
by real and virtual photons on $^3$He will be a very important test ground
to probe nuclear forces, correlated wave functions and currents. At
present the 3N  system is the only case where FSI is fully under
control (below the pion threshold) and appears therefore especially
promising to probe our  present day understanding of nuclear dynamics.

\section*{Acknowledgments}
This work has been supported by the DFG (J.G.)
and the Polish Committee for Scientific Research under Grant No. 2P03B02818.
The numerical calculations have been performed on the Cray T90 and T3E
of the NIC in J\"ulich, Germany.


\clearpage
 
\begin{thebibliography}{99}
\bibitem{ref1} W. Gl\"ockle, T.-S. H. Lee, F. Coester,
               {\it Phys, Rev.} {\bf C 33}, 709 (1986).
\bibitem{ref2} R. Machleidt, {\it Phys. Rev.} {\bf C63}, 024001 (2001).
\bibitem{ref3} R.~B.~Wiringa, V.~G.~J.~Stoks, R.~Schiavilla,
               {\it Phys. Rev.} {\bf C51}, 38 (1995).
\bibitem{ref4} V.~G.~J.~Stoks, R.~A.~M.~Klomp, C.~P.~F.~Terheggen,
               J.~J.~de Swart, {\it Phys. Rev.} {\bf C49}, 2950 (1994).
\bibitem{ref5} S. A. Coon {\it et al.},
               {\it Nucl. Phys.} {\bf A317}, 242 (1979).
\bibitem{ref6} B. S. Pudliner {\it et al.},
               {\it Phys. Rev.} {\bf C56}, 1720 (1997).
\bibitem{ref7} A. Nogga, H. Kamada, W. Gl\"ockle,
               {\it Phys. Rev. Lett.} {\bf 85}, 944 (2000).
\bibitem{ref7a} J. A. Tjon,
               {\it Phys. Lett.} {\bf B56}, 217 (1975).
\bibitem{ref8} U. van Kolck, {\it Phys. Rev.} {\bf C49}, 2932 (1994).
\bibitem{ref9} C. Ord\'o\~nez  {\it et al.},
               {\it Phys. Rev.} {\bf C53}, 2086 (1996).
\bibitem{ref10} R.~B.~Wiringa {\it et al.},
               {\it Phys. Rev.} {\bf C62}, 014001 (2000).
\bibitem{ref10a} Steven C. Pieper {\it et al.},
                {\it Phys. Rev.} {\bf C64}, 014001 (2001).
\bibitem{ref11} W. Gl\"ockle, H. Wita\l a, D. H\"uber, H. Kamada
               and J.  Golak, {\it Phys. Rep.} {\bf 274}, 107 (1996).
\bibitem{ref12} A. Kievsky  {\it et al.},
                {\it Nucl. Phys.} {\bf A607}, 402 (1996);
                {\it Phys. Rev.} {\bf C64}, 024002 (2001).
\bibitem{ref13} H. Wita{\l}a {\it et al.},
                {\it Phys. Rev.} {\bf C63}, 024007 (2001).
\bibitem{ref14} R. V. Cadman {\it et al.},
                {\it Phys. Rev. Lett.} {\bf 86}, 967 (2001).
\bibitem{ref15} W. Gl\"ockle {\it et al.},
                {\it Nucl. Phys.} {\bf A684}, 184c (2001).
\bibitem{ref16} K. Sagara {\it et al.},
                {\it Phys.Rev.} {\bf C50}, 576 (1994).
\bibitem{shimizu} H. Shimizu  {\it et al.},
                {\it Nucl. Phys.} {\bf A382}, 242 (1982).
\bibitem{ruehl} H. R\"uhl  {\it et al.},
                {\it Nucl. Phys.} {\bf A524}, 377 (1991).
\bibitem{ref17} H. Sakai  {\it et al.},
                {\it Phys. Rev. Lett.} {\bf 84}, 5288 (2000).
\bibitem{ref18} N. Sakamoto {\it et al.},
                {\it Phys. Lett.} {\bf B367}, 60 (1996).
\bibitem{ref19} H. Patberg, {\it PhD thesis,} K\"oln, 1995, unpublished. 
\bibitem{ref20} M. Allet {\it et al.},
                {\it Few Body Syst.} {\bf 20}, 27 (1996).
\bibitem{ref21} G. Rauprich  {\it et al.},
 {\it Nucl. Phys.} {\bf A535}, 313  (1991).
\bibitem{ref22} J. Zejma, {\it PhD thesis,} Krak\'ow, 1995, unpublished.
\bibitem{ref23} K. McVoy and L. van Hove,
                {\it Phys.Rev.} {\bf 125}, 1034 (1962).
\bibitem{ref24} G. Orlandini and M. Traini,
                {\it Rep. Prog. Phys.} {\bf 54}, 257 (1991).
\bibitem{ref25} J. Golak  {\it et al.},
                {\it Phys. Rev.} {\bf C52}, 1216 (1995).
\bibitem{ref26} R. Schiavilla  {\it et al.},
                {\it Phys. Rev. Lett.} {\bf 70}, 3856 (1993).
\bibitem{ref26a} D. L. Groep {\it et al.},
                {\it Phys. Rev.} {\bf C63}, 014005 (2000).
\bibitem{golak2000} J. Golak {\it et al.}, {\it Phys. Rev.} {\bf C62}, 054005 (2000).
\bibitem{ref30} W. P. Abfalterer {\it et al.},
                {\it Phys. Rev. Lett.} {\bf 81}, 57 (1998).
\bibitem{iiii} J. Golak {\it et al.}, {\it Phys. Rev.} {\bf C51}, 1638 (1995).
\end{thebibliography}
\end{document}